\documentclass[twocolumn,showpacs,prb,amsmath,amssymb]{revtex4}
\usepackage{graphicx}     
\usepackage{dcolumn}      
\usepackage{bm}           

\begin{document}

\title{Quantum antiferromagnetism and high $T_C$ superconductivity:
a close connection between the $t$-$J$ model and
the projected BCS Hamiltonian}
\author{K. Park\footnote{Future address:
School of Physics, Korea Institute for Advanced Study,
Seoul 130-722, Korea}}
\affiliation{Condensed Matter Theory Center,
Department of Physics, University of Maryland,
College Park, MD 20742-4111}
\date{\today}

\begin{abstract}
A connection between quantum antiferromagnetism
and high $T_C$ superconductivity
is theoretically investigated
by analyzing the $t$-$J$ model
and its relationships to
the Gutzwiller-projected BCS Hamiltonian.
After numerical corroboration via exact diagonalization,
it is analytically shown
that the ground state of the $t$-$J$ model at half filling
(i.e., the 2D antiferromagnetic Heisenberg model)
is entirely equivalent to the ground state of
the Gutzwiller-projected BCS Hamiltonian with strong pairing.
Combined with the high wavefunction overlap between
the ground states of the $t$-$J$ model and
the projected BCS Hamiltonian
at moderate doping,
this equivalence provides strong support for
the existence of superconductivity in the $t$-$J$ model.
The relationship between
the ground state of the projected BCS Hamiltonian
and Anderson's resonating valence bond state
(i.e., the projected BCS ground state)
is discussed.
\end{abstract}

\maketitle

\section{Introduction}
\label{intro}

There are many reasons as to why high $T_C$ superconductivity
has attracted so much attention.
Aside from the obvious (but very important) prospects for
technological applications of high $T_C$ materials,
a very salient reason is
the mysterious nature of the pairing mechanism.
Since the bare electron-electron interaction is strongly repulsive,
it seems that there are two ingredients essential to electron pairing:
(i) how to overcome the Coulomb repulsion and
(ii) how to generate an attraction between quasiparticles.

In ``low'' $T_C$ superconductivity
described well by the standard BCS theory,
the pairing mechanism is roughly as follows:
Electrons form a Fermi liquid in which
the strong Coulomb interaction is
renormalized into a screened interaction between
dressed quasiparticles
which interact very weakly;
with the strong repulsion gone,
quasiparticles can form pairs
through the exchange of phonons,
which gives rise to a time-delayed attraction
between quasiparticles.
This intuition, while valid for standard BCS superconductors,
is not correct for high $T_C$ materials.

First, in contrast to low $T_C$ superconductors
which are metallic in the normal state,
cuprates are insulators at low doping,
and thus it is not {\it a priori} clear
whether high $T_C$ superconductivity in cuprates
has anything to do with the Landau-Fermi liquid description
of the normal state in metallic systems.
In fact, this suspicion is reinforced by
many non-Fermi liquid behaviors of cuprates
including pseudogap phenomena \cite{Timusk}
and stripe excitations \cite{Stripe}.
In other words, it is not clear
how the strong Coulomb repulsion can be overcome
by standard quasiparticle screening mechanisms.

Second, there is a wealth of evidence
suggesting that pairing might be induced
by a source other than the phonon exchange.
While certainly the high energy scale of $T_C$ in the cuprates
itself is difficult to explain in terms of
the phonon exchange mechanism,
probably one of the most persuasive pieces of evidence
for a non-phonon pairing mechanism
is the destruction of superconductivity
(or the strong suppression of $T_C$)
when even a small concentration of Cu atoms
(in the Copper oxide plane)
are replaced by non-magnetic impurities
such as Zn \cite{Zn1,Zn2,Zn3}.
This is very suggestive of pairing with magnetic origin.
Also, very important in this context
is the $d$-wave symmetry of
the gap function \cite{dwave1,dwave2,dwave3,dwave4},
which is explained most naturally in terms of
pairing due to magnetic interactions.
Therefore, it seems reasonable to assume that
magnetic interactions have something to do with
high $T_C$ superconductivity.
However,
the precise relationship between
magnetism and high $T_C$ superconductivity
is not very well understood;
it is not clear exactly how the magnetic interaction
generates an attraction between electrons,
especially at low doping.
Also, from a fundamental point of view,
this attraction is very puzzling
since, after the phonon-exchange mechanism
(i.e., interaction between electrons and ions) is ruled out,
it is hard to appreciate
how the remaining microscopic interaction
could be substantially different from
the completely repulsive bare electron-electron interaction.

Despite the seemingly unrelated nature of the above two ingredients,
they are, in fact, very closely connected.
Specifically, it can be shown that
the magnetic (antiferromagnetic, to be precise) interaction is
a natural consequence of the strong repulsive interaction
between electrons at low doping,
a process known as {\it super-exchange}.
The precise mathematical derivation of super-exchange
can be carried out in the framework of model Hamiltonians.
To this end, consider
the Hubbard model with repulsive on-site interaction $U$:
\begin{eqnarray}
H_{\textrm{Hub}} = -t\sum_{\langle i,j \rangle}
(c^{\dagger}_{i\sigma}c_{j\sigma}+\textrm{H.c.})
+U\sum_i n_{i\uparrow}n_{i\downarrow}
\label{H_Hub}
\end{eqnarray}
where $\sigma=\uparrow,\downarrow$ is the spin index, and $\langle
i,j \rangle$ indicates that $i$ and $j$ are nearest neighbors. All
models studied throughout this paper are defined on the square
lattice. It has been shown \cite{Takahashi,Hirsch,GJR,MGY} that,
exactly at half filling, the Hubbard model becomes identical to
the antiferromagnetic Heisenberg model in the limit of large $U$:
\begin{eqnarray}
H_J=
J \sum_{\langle i,j \rangle}
({\bf S}_i\cdot{\bf S}_j - n_i n_j /4)
\label{H_J}
\end{eqnarray}
where $J=4t^2/U$. ${\bf S}_i$ and $n_i$ are, respectively,
the spin and the electron-number operators
at the site $i$.
Physically speaking,
the derivation of the Heisenberg model is as follows:
The zeroth order effect of $t$ for large $U$
is to completely prevent the double occupancy of any site,
thereby minimizing the large on-site interaction energy cost,
which, in turn, gives rise to a low-energy Hilbert space
in which sites are only singly occupied.
This low-energy Hilbert space, however, is hugely degenerate
since all states with single occupancy
have exactly the same energy.
Therefore, one has to investigate the next order effect of $t$.
Exactly at half filling,
there is no effect first order in $t$
because,
when it acts on the degenerate Hilbert space mentioned above,
the hopping term in Eq.(\ref{H_Hub})
always creates a doubly occupied site
which is outside the low-energy Hilbert space.
The antiferromagnetic Heisenberg model emerges
through the second-order contribution
which is the virtual hopping process
(hence, $J \propto t^2/U$).
Note that the virtual hopping process minimizes
the kinetic energy cost
in the presence of the strong on-site Coulomb repulsion.

Away from half filling with addition of holes,
the Heisenberg model generalizes to the $t$-$J$ model:
\begin{eqnarray}
H_{t\textrm{-}J}= \hat{{\cal P}}_G ( H_t + H_J ) \hat{{\cal P}}_G,
\label{H_tJ}
\end{eqnarray}
where $H_J$ is given in Eq.(\ref{H_J}) and
\begin{eqnarray}
H_t = -t\sum_{\langle i,j \rangle}
(c^{\dagger}_{i\sigma}c_{j\sigma}+\textrm{H.c.})
\label{H_t} \;.
\end{eqnarray}
The Gutzwiller projection operator, $\hat{{\cal P}}_G$,
imposes the no-double-occupancy constraint,
which handles the strong on-site interaction.
It is important to note that, to order $t^2/U$,
the rigorous derivation of the $t$-$J$ model from the Hubbard model
requires the omission of three-site hopping terms \cite{Auerbach},
which can be justified at low doping because
the hopping terms have zero matrix elements
exactly at half filling
and their contribution is
roughly proportional to the hole concentration away from half filling.
In some sense, there are two small expansion parameters
involved in deriving the $t$-$J$ model from the Hubbard model:
$t/U$ and the hole concentration, $x$.
The three-site hopping terms may become sizable
when either parameter is not small.

As shown in the above, antiferromagnetism is a natural consequence
of the strong Coulomb repulsion at low doped regimes. This is, of
course, consistent with experimental findings that there is a
well-defined long-range antiferromagnetic order (also known as
N\'{e}el order) at low doping. The next question, then, is when
and if antiferromagnetism can generate pairing; specifically,
whether the $t$-$J$ model contains superconductivity in realistic
parameter regimes (with non-zero doping). It is the goal of this
paper to provide evidence for the existence for superconductivity
in the $t$-$J$ model. To this end, we investigate the connection
between the $t$-$J$ model and the Gutzwiller-projected BCS
Hamiltonian. Note that the Gutzwiller-projected BCS Hamiltonian is
nothing but the BCS Hamiltonian in the presence of the strong
on-site repulsion.

The viewpoint advocated in this paper that antiferromagnetism and
high $T_C$ superconductivity are intimately connected is shared by
many of previous theories including the $SO(5)$ theory \cite{SO5}.
The central hypothesis of the $SO(5)$ theory is that
antiferromagnetism and superconductivity are just two different
manifestations of a single object called superspin which combines
the three-dimensional antiferromagnetic order parameter and the
two-dimensional superconducting order parameter. Apart from the
phenomenological appeal of this idea, it is crucial from the
microscopic point of view that one should be able to derive an
effective superspin model from well-known microscopic models such
as the $t-J$ model in a reliable manner. One of such attempts was
to use some form of the renormalization-group transformation
\cite{SO5}. In this paper, however, we take a different approach,
as outlined below.

This paper is organized as follows:
The precise mathematical form of the Gutzwiller-projected
BCS Hamiltonian is given in Sec. \ref{BCS_Hamiltonian}
where we discuss the physical motivation
for studying its connection to antiferromagnetism.
We also discuss
the relationship between the ground state of
the Gutzwiller-projected BCS Hamiltonian and
Anderson's resonating valence bond (RVB) state \cite{Anderson}
(i.e., the Gutzwiller-projected BCS ground state).
In Sec. \ref{numerical},
by computing the wavefunction overlap
via exact diagonalization in finite systems,
we provide evidence suggesting that
the ground state of the $t$-$J$ model
is closely connected to the ground state of
the Gutzwiller-projected BCS Hamiltonian.
In particular, we emphasize that,
within the limits of numerical accuracy,
the ground state of the antiferromagnetic Heisenberg model
(i.e., the $t$-$J$ model at half filling) is equivalent to
that of the Gutzwiller-projected BCS Hamiltonian
with strong pairing.
In fact, this equivalence can be derived analytically.
In Sec. \ref{analytic},
we provide an analytic derivation for
the equivalence between the Heisenberg model and
the Gutzwiller-projected BCS Hamiltonian with strong pairing.
We conclude in Sec. \ref{conclusion}
by discussing physical implications of
the Gutzwiller-projected BCS Hamiltonian.

\section{The Gutzwiller-projected BCS Hamiltonian}
\label{BCS_Hamiltonian}

We begin by providing a physical motivation for
studying the connection between
the $t$-$J$ model and the projected BCS Hamiltonian.
For this purpose, it is convenient to write
the pairing term of the BCS Hamiltonian in real space:
\begin{eqnarray}
H_\textrm{pair} \equiv
c^{\dagger}_{i\uparrow} c^{\dagger}_{j\downarrow}
-c^{\dagger}_{i\downarrow} c^{\dagger}_{j\uparrow}
+\textrm{H.c.}
\end{eqnarray}
which creates a resonance of
the singlet pair between the site $i$ and $j$:
$|\uparrow_i \downarrow_j\rangle-|\downarrow_i \uparrow_j\rangle$.
On the other hand,
the singlet pair is energetically preferred
by the antiferromagnetic exchange term
in the $t$-$J$ model:
${\bf S}_i \cdot {\bf S}_j$ for $J>0$.
Therefore,
despite their different appearances,
the pairing term and the antiferromagnetic exchange term
seem to have a similar physical effect;
they both prefer singlet pairs
(at least, between nearest neighbors).
This similarity serves as a motivation
to investigate whether there is
a connection between the $t$-$J$ model and
some form of the BCS Hamiltonian.
In this paper, we take
the Gutzwiller-projected BCS Hamiltonian \cite{Park}:
\begin{eqnarray}
H^{\textrm{G}}_{\textrm{BCS}} &=&
\hat{{\cal P}}_G H_{\textrm{BCS}} \hat{{\cal P}}_G
\nonumber \\
&=&
\hat{{\cal P}}_G ( H_t + H_\Delta ) \hat{{\cal P}}_G,
\label{H_GBCS}
\end{eqnarray}
where $H_t$ is given in Eq.(\ref{H_t}).
Also, the pairing term is
\begin{eqnarray}
H_{\Delta} &=& \sum_{\langle i,j \rangle}
\Delta_{ij}
\left(
c^{\dagger}_{i\uparrow} c^{\dagger}_{j\downarrow}
-c^{\dagger}_{i\downarrow} c^{\dagger}_{j\uparrow}
+\textrm{H.c.}
\right)
\label{H_Delta}
\end{eqnarray}
where we are primarily interested in
a pairing with $d$-wave symmetry, where
$\Delta_{ij}=\Delta$  if $j=i+\hat{x}$ and $-\Delta$ if $j=i+\hat{y}$.
However, we also examine
the pairing with extended $s$-wave symmetry, where
$\Delta_{ij}=\Delta$  for both $j=i+\hat{x}$ and $i+\hat{y}$,
and the Hamiltonian is denoted as $H^{\textrm{G}}_{s\textrm{BCS}}$.
Note that $H^{\textrm{G}}_{\textrm{BCS}}$ in Eq.(\ref{H_GBCS})
has a very similar structure
as $H_{t\textrm{-}J}$ in Eq.(\ref{H_tJ}).
The only change is that
$H_{\Delta}$ in $H^{\textrm{G}}_{\textrm{BCS}}$
replaces $H_J$ in $H_{t\textrm{-}J}$,
which is motivated by the possible correspondence between
the pairing term and the antiferromagnetic exchange term.
Note that $H_t + H_\Delta$ is the usual BCS Hamiltonian
in the real space representation,
and hence $H_\textrm{BCS}= H_t + H_\Delta$.

In order to provide evidence for the close connection between
the $t$-$J$ model and the Gutzwiller-projected BCS Hamiltonian,
we will pursue two approaches
with one being numerical (Sec. \ref{numerical}) and
the other being analytical (Sec. \ref{analytic}).
But, before presenting our results,
we would like to briefly discuss the relationship
between the ground state of the Gutzwiller-projected BCS Hamiltonian
and the RVB state proposed by Anderson \cite{Anderson},
which is just the Gutzwiller-projected BCS ground state:
\begin{eqnarray}
\psi_{\textrm{RVB}} = \hat{{\cal P}}_N \hat{{\cal P}}_G \psi_{\textrm{BCS}},
\label{RVB}
\end{eqnarray}
where $\hat{{\cal P}}_G$ is the previously introduced
Gutzwiller projection operator
imposing the no-double-occupancy constraint.
$\hat{{\cal P}}_N$ is the electron-number projection operator
which is necessary for constructing
states with definite electron number
from the BCS ground state, $\psi_{\textrm{BCS}}$,
that has an intrinsic fluctuation in particle number.

In general,
$\psi_{\textrm{RVB}}$ is not the exact ground state of
$H^{\textrm{G}}_{\textrm{BCS}}$
since the Gutzwiller projection does not commute with
the BCS Hamiltonian, $H_{\textrm{BCS}}$:
$[\hat{{\cal P}}_G, H_t + H_\Delta] \neq 0$.
The difference between $\psi_{\textrm{RVB}}$
and the ground state of $H^{\textrm{G}}_{\textrm{BCS}}$
is particularly severe at half filling
because, as shown in the later sections,
the ground state of $H^{\textrm{G}}_{\textrm{BCS}}$
has long-range antiferromagnetic order (N\'{e}el order)
at half filling, while $\psi_{\textrm{RVB}}$ does not.
Since the exact ground state of the $t$-$J$ model
has N\'{e}el order at half filling,
it seems that $\psi_{\textrm{RVB}}$
cannot be a good ansatz wavefunction for the $t$-$J$ model,
at least for regimes very close to half filling.
This was one of the reasons why
$\psi_{\textrm{RVB}}$ was proposed
as a candidate for the ground state of models with sufficiently
strong ``quantum frustration'' (to destroy N\'{e}el order),
examples of which include models with the next-nearest-neighbor
exchange coupling, $J'$, and antiferromagnetic models on the
triangular lattice \cite{Anderson}. Also, $\psi_{\textrm{RVB}}$
was conjectured to be a good ansatz wavefunction for the $t$-$J$
model at moderate, non-zero doping. Its validity, however, has
remained very controversial even after many years of active
research \cite{Gros,HP,PRT,Sorella,TKLee,TKLee2,ALRRTZ}.

The situation is different for
the ground state of $H^{\textrm{G}}_{\textrm{BCS}}$.
In addition to the possession of N\'{e}el order at half filling,
the ground state of $H^{\textrm{G}}_{\textrm{BCS}}$
also has a very high overlap with
the ground state of the $t$-$J$ model at moderate doping
(in fact, unity overlap at half filling),
as shown in the next section.
Therefore, the ground state of $H^{\textrm{G}}_{\textrm{BCS}}$
can be taken as a good ansatz wavefunction for
the ground state of the $t$-$J$ model at general doping.
Now, this brings up an interesting question:
how is the ground state of $H^{\textrm{G}}_{\textrm{BCS}}$
related to the RVB state at non-zero doping?
In Sec. \ref{conclusion},
it is argued that,
despite severe differences at half filling,
the RVB state is, in fact, qualitatively similar to
the ground state of $H^{\textrm{G}}_{\textrm{BCS}}$
in doped regimes sufficiently away from half filling.

\section{Numerical evidence}
\label{numerical}

In this section, we present numerical evidence for
the close connection between the $t$-$J$ model and
the Gutzwiller-projected BCS Hamiltonian.
While some numerical results are available in
a previous paper by the author \cite{Park},
more details are given in this section
along with new results.
Before we provide detailed evidence,
we would like to emphasize two important aspects of
the numerical approach used in this paper.
The first aspect pertains to the computing technique
while the second issue concerns what to compute.

First,
we use exact diagonalization as the computing technique.
Exact diagonalization offers an {\it unbiased} approach,
as opposed to other methods
with biased assumptions or guesses,
which include
the large $N$ (or $S$) expansion \cite{Auerbach,Fradkin}
and
variational Monte Carlo simulation \cite{Gros,HP,PRT,Sorella,ALRRTZ}.
Uncontrolled approximations in techniques mentioned above
require an independent verification of the involved assumptions.
In contrast,
exact diagonalization can determine
the exact ground state wavefunction without bias \cite{Dagotto}.
The problem is, however, that
its application is limited to
studies of finite systems
with relatively small spatial size.
This limitation motivates
the second aspect of the numerical approach in this paper.

The second aspect of our numerical approach concerns
what to compute.
In general,
evidence for long-range order
is provided by relevant correlation functions.
In the case of superconductivity,
the relevant correlation function is
the pairing correlation function:
\begin{eqnarray}
F_{\alpha\beta}({\bf r}-{\bf r}') =
\langle
c^{\dagger}_{\uparrow}({\bf r})
c^{\dagger}_{\downarrow}({\bf r}+{\bf \alpha})
c_{\downarrow}({\bf r}')
c_{\uparrow}({\bf r}'+{\bf \beta})
\rangle ,
\end{eqnarray}
where ${\bf \alpha}, {\bf \beta} = \hat{x}, \hat{y}$.
True off-diagonal long-range order (ODLRO)
can be claimed only when
$F_{\alpha\beta}$ remains non-zero
in the limit of large distance $|{\bf r}-{\bf r}'|$.
Unfortunately, however,
the small spatial size of finite systems
accessible via exact diagonalization makes
the distinction between true long-range order and
short-range order (present even in normal states) ambiguous.
Therefore,
a measure of pairing order
that is {\it unambiguous} even in finite system studies
is needed.

An inspiration comes from
the fractional quantum Hall effect (FQHE),
where our understanding of the subject
is significantly advanced by
the direct comparison
between ansatz wavefunctions and exact states.
It is well accepted by now that
all essential aspects of the FQHE are explained by
the composite fermion (CF) theory \cite{Jain}
which is a general theory of the FQHE
including the Laughlin state \cite{Laughlin} as a subset.
While there are various (both experimental and theoretical)
verifications of the composite fermion theory,
arguably the most significant is the amazing agreement
between the exact ground state and
the CF wavefunction:
the overlap is practically unity for
various short-range interactions including
the Coulomb interaction \cite{DasSarmaPinczuk}.
Indeed, as was crucial in establishing
the CF theory for the FQHE,
we would in turn like to achieve the same methodological clarity
for the $t$-$J$ model.

In order to give a perspective on the significance of
wavefunction overlap in finite system studies,
consider two randomly-chosen states in a Hilbert space
with $N_\textrm{basis}$ basis states.
Then, the possibility for having a large wavefunction overlap
between those two states is roughly $1/N_\textrm{basis}$.
So, roughly speaking,
if the square of the overlap
between an ansatz state and the exact state
is significantly higher than $1/N_\textrm{basis}$,
it can be argued that the ansatz state is
a good representation of the exact state.
The main finite system studied in this paper
is the $4\times4$ square lattice system,
whose Hilbert space has $10^3$ - $10^5$ basis states
depending on the number of holes
(even after translational symmetries are implemented
as reported in this paper).
Therefore, in our system,
the possibility for having a large wavefunction overlap
between two random states by chance is
roughly $0.1\%$ - $0.001\%$.
We will show in the following sections that
the square of the overlap between the ground states of
the $t$-$J$ model and the Gutzwiller-projected BCS Hamiltonian
is very close to unity at optimal parameter ranges
(typically $90\%$ - $100\%$ depending on parameters).

While it may seem straightforward at first to compute
wavefunction overlap by exactly diagonalizing
the $t$-$J$ model and the Gutzwiller-projected BCS Hamiltonian,
there is a subtle, but physically important twist
in applying exact diagonalization to the BCS Hamiltonian.
The subtlety arises from the fact that
there is a (coherent) fluctuation in particle number
due to the pairing term in $H_\Delta$
which has matrix elements mixing the Hilbert space of
$N_\textrm{e}$ and $N_\textrm{e}\pm2$ electrons.
In order to incorporate the particle-number fluctuation
into finite system studies,
we diagonalize $H^\textrm{G}_\textrm{BCS}$
in the combined Hilbert space of
$N_\textrm{e}$ and $N_\textrm{e}-2$ electrons,
which, in turn, invariably requires a careful treatment of
the chemical potential.
In essence,
the chemical potential is adjusted
so that the kinetic energy plus the chemical potential energy
of the $N_e$ particle ground state is the same as that of the
$N_e-2$ particle ground state.
Once the chemical potential is set this way,
the mixing with other particle-number sectors
such as the $N_e+2$ and $N_e\pm4$ sectors
can be shown to be negligibly small,
even if it is allowed.
For more details,
readers are referred to
discussions in the following sections.

Finally, the following notations are defined for future convenience:
$\psi^{G}_{BCS}(N_h,N_h + 2|N)$ denotes
the ground state of the Gutzwiller-projected BCS Hamiltonian
obtained from the combined Hilbert space of $N_h$ and $N_h + 2$ holes
in the system of $N$ sites.
(Note that the sum of the number of electrons and holes
equals the number of sites: $N_e + N_h = N$).
$\hat{{\cal P}}_{N_h=N_0}$ denotes
the number projection operator which projects
states onto the Hilbert space of states with $N_0$ holes,
and renormalizes the projected states.
$\psi_{t\textrm{-}J}(N_h|N)$ is the exact ground state of
the $t$-$J$ model in the Hilbert space of $N_h$ holes in $N$ sites.

Numerical evidence for the close connection
between the ground states of the $t$-$J$ model and
the Gutzwiller-projected BCS Hamiltonian
is presented as follows:
In Sec. \ref{symmetry},
the symmetry of the Gutzwiller-projected BCS Hamiltonian
with $d$-wave pairing is discussed.
Then, the wavefunction overlap between
$\psi_{t\textrm{-}J}(N_h|N)$ and
the appropriately number-projected $\psi^{G}_{BCS}(N_h,N_h + 2|N)$
is computed in Sec. \ref{undoped} for the case of undoped regime,
which is followed by similar calculations
in Sec. \ref{optimallydoped} for the optimally doped regime,
and in Sec. \ref{overdoped} for the overdoped regime.

\subsection{Symmetry}
\label{symmetry}

In general, before solving any Hamiltonian,
one has to examine the symmetry of the Hamiltonian.
Incorporating symmetry is particularly important
when two different Hamiltonians are compared
since it is possible for their ground states to have
completely different symmetries from each other.
The BCS Hamiltonian with $d$-wave pairing
is particularly tricky in this respect
because (i) it does not conserve the particle number
and (ii) it is not invariant under diagonal reflection,
i.e., $x \leftrightarrow y$
(or, equivalently, under rotation in space by $\pi/2$).
Therefore, it does not conserve parity
with respect to diagonal reflection.
Surprisingly, however,
the effects of the above two properties cancel
and can be eliminated simultaneously
by applying the number projection operator
to the ground state of the Gutzwiller-projected BCS Hamiltonian,
as explained below.
(As far as parity is concerned,
it is not important
whether the BCS Hamiltonian is Gutzwiller-projected;
the Gutzwiller projection commutes with diagonal reflection.)

First, the non-conservation of particle number is, of course,
a direct consequence of the pairing term in the BCS Hamiltonian,
$c^\dagger_\uparrow c^\dagger_\downarrow +c_\downarrow c_\uparrow$.
At first glance, the uncertainty in particle number may seem
as a theoretical artifact in the BCS Hamiltonian.
But, the coherent fluctuation in particle number is
essential for superconductivity
since it is vital to the superfluid phase coherence.
(The coherent fluctuation in superconductors
stands in sharp contrast with
the incoherent fluctuation in thermodynamic ensembles.)
On the other hand,
when a precise comparison with number eigenstates
(such as the ground state of the $t$-$J$ model) is required,
the ground state of the BCS Hamiltonian
should be projected onto
the Hilbert space of states with a definite particle number.
The number projection operator, $\hat{\cal P}_N$,
performs this task.

Second, the non-conservation of parity
with respect to diagonal reflection is due to
the sign difference in $d$-wave pairing amplitudes
between the $x$ and $y$ direction.
To be more precise,
the non-conservation of parity
is mathematically expressed as follows:
\begin{eqnarray}
\hat{\cal R}_\textrm{d}
H^{\textrm{G}}_{\textrm{BCS}}(t,\Delta)
\hat{\cal R}_\textrm{d}
=
H^{\textrm{G}}_{\textrm{BCS}}(t,-\Delta)
\label{parity1}
\end{eqnarray}
where $\hat{\cal R}_\textrm{d}$ is the diagonal reflection operator
and $H^{\textrm{G}}_{\textrm{BCS}}(t,\Delta)$
is the Gutzwiller-projected BCS Hamiltonian
given in Eq.(\ref{H_GBCS})
with the dependence on $t$ and $\Delta$ shown explicitly.
Note the sign change of $\Delta$ in $H^{\textrm{G}}_{\textrm{BCS}}$
in the right-hand side of Eq.(\ref{parity1}).
Now, for convenience,
Eq.(\ref{parity1}) is re-written as follows:
\begin{eqnarray}
\hat{\cal R}_\textrm{d}
H^{\textrm{G}}_{\textrm{BCS}}(t,\Delta)
=
H^{\textrm{G}}_{\textrm{BCS}}(t,-\Delta)
\hat{\cal R}_\textrm{d}
\label{parity2}
\end{eqnarray}
since $\hat{\cal R}^2_\textrm{d}=1$.
What we want to prove in this section is that
a number-projected eigenstate of $H^{\textrm{G}}_{\textrm{BCS}}$
is also a parity eigenstate.

We begin by defining $|\Psi\rangle$ as
an eigenstate of $H^{\textrm{G}}_{\textrm{BCS}}(t,\Delta)$:
\begin{eqnarray}
H^{\textrm{G}}_{\textrm{BCS}}(t,\Delta) |\Psi\rangle
= E |\Psi\rangle .
\label{eigen1}
\end{eqnarray}
Also, for convenience,
let us write $|\Psi\rangle$ in vector form
showing amplitudes in each individual particle-number sector.
That is,
\begin{eqnarray}
\langle \Psi | =
\left(
\cdots,C^{\alpha}_{N_e},\cdots,C^{\beta}_{N_e+2},\cdots
\right) ,
\end{eqnarray}
where $C^{\alpha}_{N_e}$ is the amplitude for the $\alpha$-th basis element
in the $N_e$ particle Hilbert space. Similarly,
$C^{\beta}_{N_e+2}$ is the amplitude for the $\beta$-th basis element
in the $N_e+2$ particle Hilbert space.

Now, it is not too difficult to show that
\begin{eqnarray}
H^{\textrm{G}}_{\textrm{BCS}}(t,-\Delta) |\tilde{\Psi}\rangle
= E |\tilde{\Psi}\rangle ,
\label{eigen2}
\end{eqnarray}
where $E$ is the same energy as in Eq.(\ref{eigen1}) and
\begin{eqnarray}
\langle\tilde{\Psi}| =
\left(
\cdots,[-1]^\frac{N_e}{2} C^{\alpha}_{N_e},\cdots,[-1]^{\frac{N_e+2}{2}}
C^{\beta}_{N_e+2},\cdots
\right) ,
\label{eigen2_state}
\end{eqnarray}
where $N_e$ is restricted to even numbers
since we are interested only in paired states.
What accounts for
the form of Eq.(\ref{eigen2_state}) is that
the pairing term, $H_\Delta$, always mixes
the Hilbert spaces with particle numbers differing by two.
Attaching the relative sign according to the particle number
(i.e., $[-1]^\frac{N_e}{2}$ and $[-1]^\frac{N_e+2}{2}$ for
the $N_e$ and $N_e+2$ Hilbert space, respectively)
is tantamount to changing the sign of $\Delta$ in $H_\Delta$.
Note that a relative sign difference
between distinct Hilbert spaces
does not alter the hopping term
since $H_t$ does not mix Hilbert spaces
with different particle numbers.

Now, let $\hat{\cal R}_d$ act on both sides of Eq.(\ref{eigen1}):
\begin{eqnarray}
\hat{\cal R}_d H^{\textrm{G}}_{\textrm{BCS}}(t,\Delta) |\Psi\rangle
= E \hat{\cal R}_d |\Psi\rangle ,
\end{eqnarray}
which, with aids of Eq.(\ref{parity2}), becomes
\begin{eqnarray}
H^{\textrm{G}}_{\textrm{BCS}}(t,-\Delta) \hat{\cal R}_d |\Psi\rangle
= E \hat{\cal R}_d |\Psi\rangle .
\label{eigen3}
\end{eqnarray}
Then, by comparing Eq.(\ref{eigen2}) and (\ref{eigen3}),
one is able to conclude that
\begin{eqnarray}
\hat{\cal R}_d |\Psi\rangle = \lambda |\tilde{\Psi}\rangle ,
\end{eqnarray}
where $\lambda$ is a constant which is almost
(though not quite yet) an eigenvalue.
After acting on both sides by $\hat{\cal P}_{N=N_e}$,
the above equation becomes
\begin{eqnarray}
\hat{\cal R}_d \hat{\cal P}_{N=N_e} |\Psi\rangle =
\lambda \hat{\cal P}_{N=N_e} |\tilde{\Psi}\rangle .
\label{parity_conclusion1}
\end{eqnarray}
(Note that $\hat{\cal P}_N$ commutes with $\hat{\cal R}_d$.
Also, there is no loss of generality in choosing $N=N_e$.)
Moreover, since
\begin{eqnarray}
\hat{\cal P}_{N=N_e} |\tilde{\Psi}\rangle
= (-1)^\frac{N_e}{2} \hat{\cal P}_{N=N_e} |\Psi\rangle ,
\end{eqnarray}
Eq.(\ref{parity_conclusion1}) becomes
\begin{eqnarray}
\hat{\cal R}_d \hat{\cal P}_{N=N_e} |\Psi\rangle =
\lambda' \hat{\cal P}_{N=N_e} |\Psi\rangle ,
\label{parity_conclusion2}
\end{eqnarray}
where $\lambda' = (-1)^\frac{N_e}{2} \lambda$.
Consequently,
$\hat{\cal P}_N |\Psi\rangle$ is an eigenstate of
the parity with respect to the diagonal reflection.

Until now, only the spatial symmetry has been investigated.
(Note that $H^{\textrm{G}}_{\textrm{BCS}}$ is
invariant under the spatial translation, and therefore
linear momentum is conserved.)
Now, we would like to briefly consider spin rotation symmetry.
The hopping term is obviously invariant under spin rotation.
In addition,
the pairing term can be also shown to be invariant under the spin rotation:
$[H_\Delta,{\bf S}_{\textrm{tot}}]=0$
where the total spin is ${\bf S}_{\textrm{tot}}=\sum_i {\bf S}_i$.
The essential physics of
the spin-rotational invariance of the pairing term
is the fact that
$H_\Delta$ concerns only spin-singlet pairs,
which are rotationally invariant.

\subsection{Undoped regime (half filling)}
\label{undoped}

In this section, we provide numerical evidence indicating that,
at half filling,
the ground state of the Gutzwiller-projected BCS Hamiltonian
is actually equivalent to the ground state of the $t$-$J$ model
in the limit of strong pairing, i.e., $\Delta/t\rightarrow\infty$.
To this end,
the wavefunction overlap between the two ground states
is computed as a function of $\Delta/t$
via a modified Lanczos method for exact diagonalization.
Note that the ground state of the $t$-$J$ model
is uniquely determined at half filling
without any dependence on $J/t$,
while the ground state of the Gutzwiller-projected BCS Hamiltonian
depends on $\Delta/t$.
The reason for the former is that, at half filling,
the $t$-$J$ model becomes the antiferromagnetic Heisenberg model
where $J$ is just an overall scale factor.

As mentioned previously,
the main finite system studied in this paper is
the $4\times4$ square lattice system
with periodic boundary conditions.
The $4\times4$ system is one of the most studied
systems in numerical treatments \cite{Dagotto}
because it is accessible
via exact diagonalization,
yet large enough to contain
essential many-body correlations.
We have checked that our results for the $t$-$J$ model
are in complete agreement with previous numerical studies \cite{DagottoPRB}
for all available cases.
Using notations defined earlier in Sec. \ref{numerical},
the ground state of the $t$-$J$ model is denoted
as $\psi_{t\textrm{-}J}(0|16)$ for half filling:
0 holes (16 electrons) in the $4\times4$ system.
Also, the ground state of the Gutzwiller-projected BCS Hamiltonian
at half filling
is obtained by applying $\hat{{\cal P}}_{N_h=0}$
to $\psi^{G}_{BCS}(0,2|16)$.
Note that $\psi^{G}_{BCS}(0,2|16)$ is
the ground state of the combined Hilbert space of 0 and 2 holes
(16 and 14 electrons, respectively) in the $4\times4$ system.

\begin{figure}
\includegraphics[width=1.8in,angle=-90]{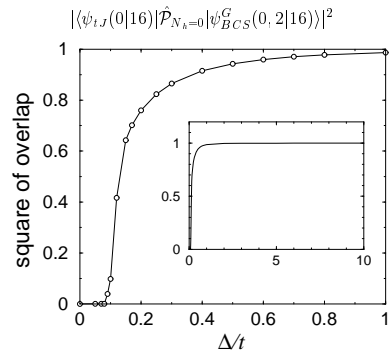}
\caption{Square of the overlap between
the ground states of the $t$-$J$ model at half filling
(i.e., the 2D antiferromagnetic Heisenberg model)
and the projected BCS Hamiltonian.
Exact diagonalization is performed
in the 4$\times$4 system with 0 holes
(undoped regime).
Inset: plotted is the square of the overlap
in an extended range of $\Delta/t$ up to 10
showing quick saturation of the overlap to unity.
\label{fig1}}
\end{figure}

At half filling, the overlap between the ground states of the
$t$-$J$ model and the Gutzwiller-projected BCS Hamiltonian is
given by $\langle\psi_{t\textrm{-}J}(0|16)| \hat{{\cal P}}_{N_h=0}
|\psi^{G}_{BCS}(0,2|16)\rangle$, the square of which is plotted in
Fig. \ref{fig1} as a function of $\Delta/t$. As one can see from
Fig. \ref{fig1}, the overlap is very high even for $\Delta/t
\simeq 1$, and saturates very quickly to unity as $\Delta/t$
increases further. For sufficiently large values of $\Delta/t$,
the overlap is indistinguishable from unity to within numerical
accuracy (We have actually studied $\Delta/t$ values ranging as
high as 1000). Therefore, as far as our finite system is
concerned, the ground state of the Gutzwiller-projected BCS
Hamiltonian is exactly identical to the ground state of the
$t$-$J$ model in the limit of strong pairing, i.e.,
$\Delta/t\rightarrow\infty$. In view of the fact that there are
roughly $10^3$ basis states in the Hilbert space for the
$4\times4$ system at half filling even after translational
symmetries are implemented, the high overlap is particularly
salient. We have also checked that the above result also holds in
the 4-site ($2\times2$) and the 10-site
($\sqrt{10}\times\sqrt{10}$) systems.
It should be emphasized that
the equivalence between the ground state of the $t$-$J$ model and
that of the projected BCS Hamiltonian with strong pairing does not
necessarily mean strong superconductivity since, despite strong
pairing, there is little charge fluctuation near half filling, and
therefore little phase coherence.

The equivalence between the ground states of
the projected BCS Hamiltonian and
the $t$-$J$ model at half filling
has a very important physical implication
for the origin of high $T_C$ superconductivity
supporting a long-standing conjecture \cite{Anderson}.
The conjecture is that
electrons are already paired at half filling,
and therefore form a condensate.
However, electrons cannot superconduct
(or, for that matter, even conduct)
at half filling because there is no room for them to travel.
But, away from half filling,
it seems natural that
removing some fraction of electrons (i.e., doping)
may trigger superconductivity
by mobilizing electrons
(and also causing charge fluctuations).
In support of this idea,
we show in Sec. \ref{optimallydoped} that
the overlap between the ground states of
the projected BCS Hamiltonian and the $t$-$J$ model
remains very high even at non-zero doping
for reasonable parameter ranges.

\subsection{Optimally doped regime}
\label{optimallydoped}

In this section,
we present numerical results for
the wavefunction overlap between
the ground states of
the projected BCS Hamiltonian and the $t$-$J$ model
at a moderate, non-zero doping.
Specifically,
we study the situation
in which there are 2 holes in the $4\times4$ system,
which roughly corresponds to the optimally doped regime.
Using notations defined earlier,
the overlap is given by
$\langle\psi_{t\textrm{-}J}(2|16)|
\hat{{\cal P}}_{N_h=2}
|\psi^{G}_{BCS}(0,2|16)\rangle$
for the Hilbert space with 2 holes.
An alternative representation of the overlap
for this Hilbert space can be obtained from
$\langle\psi_{t\textrm{-}J}(2|16)|
\hat{{\cal P}}_{N_h=2}
|\psi^{G}_{BCS}(2,4|16)\rangle$
which, however,
leads to essentially the same conclusion,
as shown later in this section.

\begin{figure}
\includegraphics[width=2.4in]{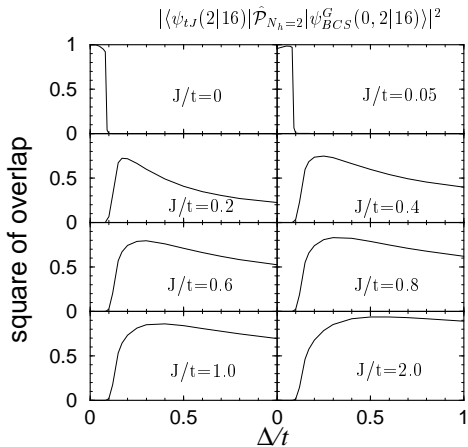}
\caption{
Square of the overlap between
the ground states of the $t$-$J$ model
and the projected BCS Hamiltonian
in the 4$\times$4 system with 2 holes
(which roughly corresponds to
the optimally doped regime).
\label{fig2}}
\end{figure}
\begin{figure}
\includegraphics[width=2.7in]{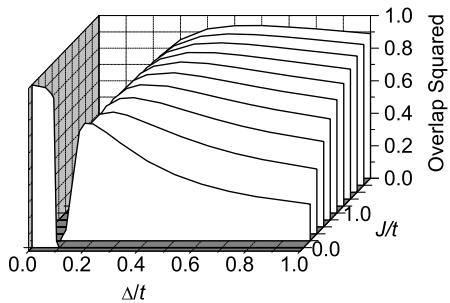}
\caption{Three-dimensional plot for the square of the overlap
as a function of $\Delta/t$ and $J/t$
in the 4$\times$4 system with 2 holes.
While this plot is essentially the same
as Fig.\ref{fig2},
the existence of
two distinct regions of high overlap
is more easily visualized.
\label{fig3}}
\end{figure}
\mbox{Fig. \ref{fig2}} displays
square of the overlap defined by
$|\langle\psi_{t\textrm{-}J}(2|16)|
\hat{{\cal P}}_{N_h=2}
|\psi^{G}_{BCS}(0,2|16)\rangle|^2$.
Note that the overlap now is a function of two parameters:
$\Delta/t$ for the projected BCS Hamiltonian and
$J/t$ for the $t$-$J$ model.
For better visualization,
a three-dimensional plot of the overlap
is also given in \mbox{Fig. \ref{fig3}}.
As shown in \mbox{Fig. \ref{fig2}} and
\mbox{Fig. \ref{fig3}}
there are two distinct regimes of high overlap:
a weak-coupling regime
($J/t \lesssim 0.08$ and $\Delta/t \lesssim 0.1$)
and a strong-coupling regime
($J/t \gtrsim  0.08$ and $\Delta/t \gtrsim 0.1$).
These two regimes are qualitatively different
in the sense that
the symmetry of the ground states
with respect to a spatial rotation by $\pi/2$
changes from $s$-wave to $d$-wave
at the boundary $J/t \simeq 0.08$
for the ground state of the $t$-$J$ model,
and at the boundary $\Delta/t \simeq 0.1$
for the ground state of the projected BCS Hamiltonian.
Note that, in a similar fashion
for the case of parity with respect to diagonal reflection,
the angular momentum associated with spatial rotation by $\pi/2$
can be shown
to be conserved in the BCS Hamiltonian with $d$-wave pairing.
Note also that, because of the symmetry change,
the overlap in the regime defined by
$J/t \gtrsim 0.08$ and $\Delta/t \lesssim 0.1$ is precisely zero.

The high overlap in the weak-coupling regime is rather trivial
because, in this regime,
both the projected BCS Hamiltonian and
the $t$-$J$ model Hamiltonian
are reduced basically to the same Hamiltonian
which is just the hopping Hamiltonian, $H_t$,
with the Gutzillwer projection;
hence, we regard the equivalence at small $J/t$ and $\Delta/t$
as a self-consistency check
for the techniques used in this paper.
The high overlap in the strong-coupling regime,
on the other hand,
cannot be trivially explained.
As seen in \mbox{Fig. \ref{fig2}},
the maximum value of the overlap (as a function of $\Delta/t$)
approaches unity as $J/t$ increases,
which can be also seen in \mbox{Fig. \ref{fig3}}
in the form of a rising ridge as $J/t$ increases.
The high overlap in the strong-coupling regime
is therefore not accidental,
but rather is connected to the unity overlap
in the limit of strong coupling,
which is in turn due to an intrinsic connection
between antiferromagnetism and superconductivity.

\begin{figure}
\includegraphics[width=2.4in]{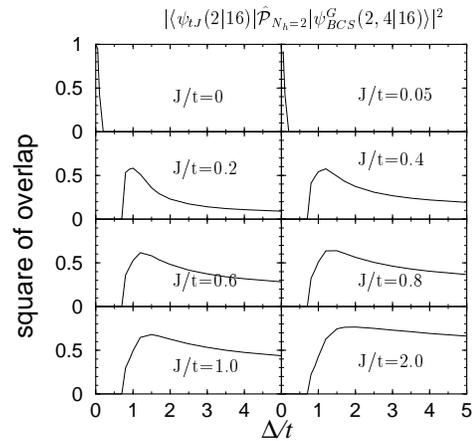}
\caption{Square of the overlap
for the 4$\times$4 system with 2 holes,
which is in the same regime as Fig.\ref{fig2}.
But, the ground state of the projected BCS Hamiltonian
in this plot,
$\hat{\cal P}_{N_h=2}|\psi^{G}_{BCS}(2,4|16)\rangle$,
is obtained from the combined Hilbert space of
2 and 4 holes in the $4\times4$ system,
as opposed to the combined Hilbert space of
0 and 2 holes in Fig.\ref{fig2}.
It is important to note that
qualitative features of this plot are
basically identical to \mbox{Fig. \ref{fig2}},
while there are some quantitative differences.
\label{fig4}}
\end{figure}
Finally, as mentioned before,
we study the overlap defined by
$\langle\psi_{t\textrm{-}J}(2|16)|
\hat{{\cal P}}_{N_h=2}
|\psi^{G}_{BCS}(2,4|16)\rangle$.
This alternative definition is also valid because
the ground state of the projected BCS Hamiltonian
can be obtained in two different ways:
one can apply $\hat{{\cal P}}_{N_h=2}$ either
to the ground state of the combined Hilbert space of 0 and 2 holes,
or to that of the combined Hilbert space of 2 and 4 holes.
While, strictly speaking,
these two definitions become identical
only in the thermodynamic limit,
it would be an assuring self-consistency check of our approach
if they produce similar results
even in the finite system studies that we study.
We find that this is indeed the case.
\mbox{Fig. \ref{fig4}}
depicts
$|\langle\psi_{t\textrm{-}J}(2|16)|
\hat{{\cal P}}_{N_h=2}
|\psi^{G}_{BCS}(2,4|16)\rangle|^2$
showing that the essential features are
basically identical to those in \mbox{Fig. \ref{fig2}},
while there are some minor quantitative differences.

\subsection{Overdoped regime}
\label{overdoped}

\begin{figure}
\includegraphics[width=1.8in]{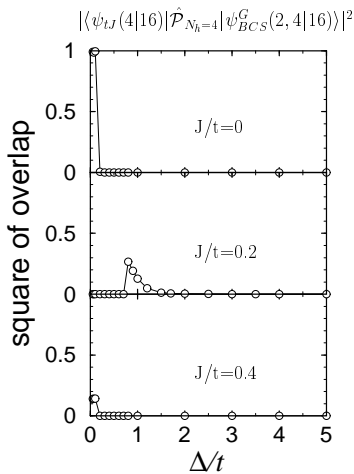}
\caption{Square of the overlap between
the ground states of the $t$-$J$ model
and the projected BCS Hamiltonian
in the 4$\times$4 system with 4 holes
(which roughly corresponds to
an overdoped regime).
\label{fig5}}
\end{figure}

We now present numerical results
for the case of 4 holes in the $4\times4$ system,
which roughly corresponds to the overdoped regime.
In this regime,
it is not strictly appropriate to regard
the $t$-$J$ model as the large-$U$ limit of the Hubbard model
due to the omission of three-site hopping terms
mentioned in the introduction.
Experimentally, however,
superconductivity is weakened and eventually destroyed
as the hole concentration increases.
Hence, it is natural to expect that
the overlap between the ground states of
the projected BCS Hamiltonian and the $t$-$J$ model
becomes small in this regime.
We show in \mbox{Fig. \ref{fig5}}
that this trend does indeed occur.

\mbox{Fig. \ref{fig5}}
displays
$|\langle\psi_{t\textrm{-}J}(4|16)|
\hat{{\cal P}}_{N_h=4}
|\psi^{G}_{BCS}(2,4|16)\rangle|^2$,
which shows that
the overlap is, in general, negligibly small
except for the trivial case of $J/t=0$ and $\Delta/t=0$.
Therefore, in the overdoped regime,
the ground state of the projected BCS Hamiltonian
is no longer a good representation of
the ground state of the $t$-$J$ model.

\section{Analytic derivation of the equivalence at half filling}
\label{analytic}

It has been shown numerically in the previous section
that, at half filling,
the ground state of the Gutzwiller-projected BCS Hamiltonian
is equivalent to the exact ground state of the $t$-$J$ model
in the limit of strong pairing, i.e., $\Delta/t \rightarrow \infty$.
While the numerical evidence is quite strong
(the overlap between the two ground states
is indistinguishable from unity to within numerical accuracy),
questions regarding the validity of finite-system studies
linger: especially, whether the overlap is actually equal to unity,
or just very close to unity.
If only the latter is true,
the overlap will diminish and
will eventually vanish
as the system size increases.

We emphasize that,
even if it is true
that the overlap does vanish in the thermodynamic limit,
the high wavefunction overlap for the finite system
still provides
strong support for the ansatz wavefunction
by demonstrating that
it contains the correct physics.
This situation is, in fact, very similar to
what happens in the FQHE:
the overlap between the Laughlin wavefunction
(the CF wavefunction, in general) and
the exact ground state of the Coulomb interaction
decreases as the system size increases,
and eventually vanishes in the thermodynamic limit.
The vanishing overlap in the thermodynamic limit
is an inevitable consequence of the fact that
the two ground states are not precisely identical.
For example,
while the Laughlin state
at the lowest-Landau-level filling factor $\nu=1/3$
is the exact ground state of the short-range interaction given by
$\nabla^2 \delta({\bf r})$ \cite{Haldane,TK},
it is still an approximation
for the exact ground state of the Coulomb interaction
(relevant for experiments)
albeit an extremely good one.

In this paper,
it will be shown that the overlap at half filling
is actually unity in the strong-pairing limit;
in other words,
we will prove that, at half filling,
the ground state of
the Gutzwiller-projected BCS Hamiltonian with strong pairing
is identical to the ground state of the Heisenberg model.
We will also prove a stronger statement that
the two Hamiltonians do not merely share the same ground state,
but also have in common the same low-energy physics.
Note that
the Hamiltonian for the Heisenberg model is $H_J$ in \mbox{Eq. (\ref{H_J})}
and the Gutzwiller-projected BCS Hamiltonian with strong pairing
is given by
$H^\textrm{G}_\textrm{BCS}$ in \mbox{Eq. (\ref{H_GBCS})} with $t=0$,
which is nothing but $H_\Delta$ with the Gutzwiller projection:
\begin{eqnarray}
H^\textrm{G}_\Delta =
\hat{{\cal P}}_G
\Big[
\sum_{\langle i,j \rangle} \Delta_{ij}
(c^{\dagger}_{i\uparrow}c^{\dagger}_{j\downarrow}
-c^{\dagger}_{i\downarrow}c^{\dagger}_{j\uparrow}
+\textrm{H.c.})
\Big]
\hat{{\cal P}}_G .
\label{H_GDelta}
\end{eqnarray}
Note that the strong-coupling limit,
i.e. the large-$\Delta/t$ limit,
is equivalent to the situation in which $t=0$
since $t/\Delta=0$ is not a singular point.

\subsection{Gutzwiller projection as the large-$U$ limit
and Lieb's theorem}

We begin our analytic derivation for the equivalence at half filling
by writing the BCS Hamiltonian
with finite, repulsive on-site interaction $U$:
\begin{eqnarray}
H_{\textrm{BCS}+U}(t,\Delta_0)
= H_t + H_{\Delta_0} +U\sum_i n_{i\uparrow}n_{i\downarrow},
\end{eqnarray}
where $H_t$ is defined in \mbox{Eq. (\ref{H_t})}.
It is very important to note that
$H_{\Delta_0}$ is given in \mbox{Eq. (\ref{H_Delta})}
with a {\it bare} pairing amplitude, $\Delta_0$,
instead of a fully renormalized pairing amplitude, $\Delta$.
The relationship between $\Delta_0$ and $\Delta$
should become clear at the end of the analytic derivation:
we mention in advance, however, that
$\Delta \propto \Delta_0^2/U$.

Our study of $H_{\textrm{BCS}+U}$
is motivated by the fact that, in the large-$U$ limit,
$H_{\textrm{BCS}+U}$ reduces to
the Gutzwiller-projected BCS Hamiltonian.
In addition,
treating the Gutzwiller projection as the large-$U$ limit
instead of working directly with the projection
facilitates the analysis.
Motivated by numerical studies given in the previous section,
we are particularly interested in $H_{\textrm{BCS}+U}$
with strong pairing:
\begin{eqnarray}
H_{\Delta_0+U} &=& H_{\textrm{BCS}+U}(t=0,\Delta_0)
\nonumber \\
&=& H_{\Delta_0}  +U\sum_i n_{i\uparrow}n_{i\downarrow}
\end{eqnarray}
which, in the large-$U$ limit,
becomes $H^\textrm{G}_\Delta$ in \mbox{Eq. (\ref{H_GDelta})}
What we strive to prove is that, in the limit of large $U/\Delta_0$,
$H_{\Delta_0+U}$ has exactly the same low-energy physics
(i.e., the same ground state and the same low-energy excitations)
as the Hubbard model, $H_\textrm{Hub}$, in the limit of large $U/t$.
This, in turn, means
that the low-energy physics is identical for both
$H^\textrm{G}_\Delta$ and $H_J$
since $H_\textrm{Hub}$ becomes $H_J$ in the large-$U/t$ limit.

It is important to note
a subtle, but very crucial difference
between the large-$U$ behavior of the Hubbard model
and that of the strong-pairing Gutzwiller-projected BCS Hamiltonian.
To appreciate this distinction,
let us re-write the Hamiltonian for the Hubbard model
as follows:
\begin{eqnarray}
H_{\textrm{Hub}} &=& H_t +U\sum_i n_{i\uparrow}n_{i\downarrow}
\nonumber \\
&=& H_{\textrm{BCS}+U}(t,\Delta_0=0),
\end{eqnarray}
which shows that
$H_\textrm{Hub}$ is a special case of
$H_{\textrm{BCS}+U}$ with $\Delta_0=0$,
while $H_{\Delta+U}$ was also a special case of
$H_{\textrm{BCS}+U}$ but with $t=0$.
Hence, there is a parallel between
$H_\textrm{Hub}$ and $H_{\Delta_0+U}$.
Specifically,
the equivalence between the low-energy physics
of $H^\textrm{G}_\Delta$ and $H_J$
can be recast as
the equivalence between the low-energy physics
of two different parameter points
(i.e, between those of $t/\Delta_0=0$ and $t/\Delta_0=\infty$)
of the same Hamiltonian, $H_{\textrm{BCS}+U}(t,\Delta_0)$,
in the large-$U$ limit.

In fact, it has been shown by Affleck {\it et al.} \cite{Affleck}
that two apparently very different mean-field solutions of
the half-filled Hubbard model are actually equivalent.
The first mean-field solution is
considered by Baskaran, Zou, and Anderson \cite{BZA}
as well as
by Ruckenstein, Hirschfeld, and Appel \cite{RHA}
and later by Kotliar \cite{Kotliar},
who performed a quadratic factorization of $H_J$
assuming that
$\langle c_{i\uparrow} c_{j\downarrow} \rangle \neq 0$
and $\langle c^\dagger_{i\sigma} c_{j\sigma} \rangle = 0$
for nearest neighbor $i$ and $j$.
On the other hand,
the second mean-field solution,
considered by Affleck and Marston \cite{AM},
assumes the opposite situation:
$\langle c_{i\uparrow} c_{j\downarrow} \rangle = 0$
and $\langle c^\dagger_{i\sigma} c_{j\sigma} \rangle \neq 0$.
It thus seems that
the mean-field solution with
$\langle c_{i\uparrow} c_{j\downarrow} \rangle \neq 0$
and $\langle c^\dagger_{i\sigma} c_{j\sigma} \rangle = 0$,
which corresponds to $H_{\textrm{BCS}+U}$ with $t=0$,
is equivalent to that with
$\langle c_{i\uparrow} c_{j\downarrow} \rangle = 0$
and $\langle c^\dagger_{i\sigma} c_{j\sigma} \rangle \neq 0$,
which corresponds to $H_{\textrm{BCS}+U}$ with $\Delta_0=0$.
In this paper,
we prove that
a stronger equivalence,
not just between mean-field solutions,
but also
between the exact ground state of $H_{\textrm{BCS}+U}$
with $t=0$ and with $\Delta_0=0$.
We, however, emphasize that,
despite their success in mean-field theory,
approaches based on unitary transformations
are not applicable in the exact treatment
due to the fundamental difference in
the large-$U$ behavior of
$H_\textrm{Hub}$ and $H_{\Delta_0+U}$,
which is discussed in greater detail below.

Consider the effect of a large $U$ in $H_\textrm{Hub}$.
In this situation,
the zeroth-order effect of $H_t$ corresponds to
keeping only the on-site repulsion term,
in which case the ground state energy
is exactly the same for arbitrary spin configurations
as long as there is a single electron per site.
Therefore, there is a huge $2^N$ degeneracy
in the low-energy Hilbert space
which is, in fact, the Gutzwiller-projected space.
Now, let us investigate the next order
effect in the Gutzwiller-projected space.
The first-order effect of $H_t$
does not contribute to
the Gutzwiller-projected space
since, at half filling,
the hopping term always creates a doubly occupied site
taking the state beyond the Gutzwiller-projected space.
In other words, exactly at half filling,
\begin{eqnarray}
\hat{\cal P}_\textrm{G} H_t \hat{\cal P}_\textrm{G} = 0 .
\end{eqnarray}
The antiferromagnetic Heisenberg model, $H_J$,
arises from
the second-order virtual hopping processes.

Next, we consider the effect of a large $U$ in $H_{\Delta_0+U}$.
The zeroth-order effect of $H_{\Delta_0}$
is, of course, the same as that of $H_t$ in $H_\textrm{Hub}$:
the Gutzwiller-projected space
is the low-energy Hilbert space.
In contrast to $H_\textrm{Hub}$, however,
there is a non-zero first-order effect of pairing term:
\begin{eqnarray}
H^\textrm{G}_\Delta=
\hat{\cal P}_\textrm{G} H_\Delta \hat{\cal P}_\textrm{G} ,
\end{eqnarray}
where $\Delta$ is renormalized from the bare value $\Delta_0$.
Physically speaking,
the strong-pairing BCS Hamiltonian sets up
the pairing resonance directly instead of by virtual processes.
Therefore,
it is not {\it a priori} clear why or if
$H^\textrm{G}_\Delta$ has the same low-energy physics as $H_J$.
In fact, the above real v.s. virtual contrast
suggests that, if it exists,
the equivalence between the low-energy physics
of $H^\textrm{G}_\Delta$ and $H_J$
cannot be derived via a simple unitary transformation of the Hamiltonian.
Instead, the equivalence must be connected to
intricate physics.

An essential point is that
the large-$U$ behavior of $H_{\Delta_0+U}$ and $H_\textrm{Hub}$
can be dealt with systematically in our approach
where we analyze finite-$U$ models and
take the large-$U$ limit as the Gutzwiller projection.
There is no singular behavior in the large-$U$ limit of $H_{\Delta_0+U}$
(because $H^\textrm{G}_\Delta$ is non-zero).
However,
it is not obvious whether the ground state of
the antiferromagnetic Heisenberg model is
adiabatically connected to those of the Hubbard model
with finite $U$.
In fact, it can be proven by Nagaoka's theorem \cite{Nagaoka}
that, infinitesimally away from half filling,
the ground state of the Gutzwiller-projected hopping term,
$\hat{\cal P}_\textrm{G} H_t \hat{\cal P}_\textrm{G}$,
(which, in a na\"{i}ve sense, is
the large-$U$ limit of the Hubbard model)
is ferromagnetic rather than antiferromagnetic.
Therefore, one should be careful
in how one takes this limit.
Fortunately,
there is a theorem by Lieb \cite{Lieb}
showing that the ground state of the Hubbard model
is uniquely determined at {\it any positive} $U$
and, as a corollary,
the ground state in the large-$U$ limit is
adiabatically connected to those of finite $U$.
Therefore,
the ground state of
the antiferromagnetic Heisenberg model is
adiabatically connected to ground states of the Hubbard model
for which $U$ is finite.
(We assume that this connection is also valid
for low-energy excitations.)

We now present our analytic derivation step by step
beginning with an outline summarizing each step:
(i) We begin by separating $H_{\Delta_0+U}$ and
$H_{\textrm{Hub}}$
into two parts:
the saddle-point Hamiltonian
(${\cal H}_{\Delta_0+U}$ and ${\cal H}_{\textrm{Hub}}$,
respectively)
and the remaining Hamiltonian incorporating
fluctuations around the saddle-point solution
($\delta {\cal H}_{\Delta_0+U}$ and
$\delta {\cal H}_{\textrm{Hub}}$,
respectively).
The saddle-point Hamiltonian is chosen
so as to capture a possible singular effect of
the spin-density-wave (SDW) instability at ($\pi,\pi$),
which is inherent in $H_{\Delta_0}$ and $H_t$
in the presence of repulsive on-site interactions
with arbitrary strength.

(ii) The ground states of
${\cal H}_{\Delta_0+U}$ and ${\cal H}_{\textrm{Hub}}$
are obtained for a general $U$.
It is shown that, in both cases,
the saddle-point ground state is separated from other excited states
by an energy gap proportional to $U$ when $U$ is large.
It is also shown that these two saddle-point ground states
become identical in the large-$U$ limit.
We denote this saddle-point ground state in the large-$U$ limit
as $|\psi_0\rangle$.

(iii) In the large-$U$ limit,
the low-energy Hilbert space for the full Hamiltonian
(including the saddle-point and fluctuation part)
is composed only of states that are connected to
$|\psi_0\rangle$ via rigid spin rotations, $\{ {\bf R}_i \}$.
We denote these states as $\{|\psi_i\rangle\}$.
Then, we show that in the large-$U$ limit,
all matrix elements of
$\delta {\cal H}_{\Delta_0+U}$ and $\delta {\cal H}_{\textrm{Hub}}$
are precisely the same
in the low-energy Hilbert space
with the same being true for
the matrix elements of the two saddle-point Hamiltonians.

\begin{figure}
\includegraphics[width=3in]{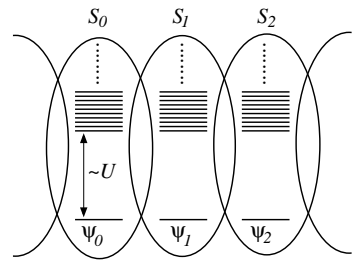}
\caption{Schematic diagram for the structure of
the Hilbert space expanded by saddle-point basis states.
The vertical separation between short lines
denotes the energy difference
between saddle-point states.
A new sub-Hilbert-space $S_i$ is obtained by
applying a rigid spin rotation ${\bf R}_i$ to
every state in $S_0$.
Without a loss of generality,
one can choose $S_0$ to be the Hilbert space
with spin quantization axis
aligned with the long-range spin order.
$\psi_i$ denotes the lowest-energy state of $S_i$.
Note that the above sub-Hilbert-spaces are linearly independent,
but not orthogonal to each other
(hence, overlapping ellipses).
In the limit of large $U$,
$\psi_i$'s form an over-complete basis set for
the Hilbert space at half filling,
which is conceptually very similar to
the coherent-state basis set for a spin representation.
\label{fig6}}
\end{figure}

(iv) Having shown that
all matrix elements in the low-energy Hilbert space
are precisely the same in the large-$U$ limit,
we argue that
the strong-pairing Gutzwiller-projected BCS Hamiltonian
and the 2D antiferromagnetic Heisenberg model
have identical low-energy physics.
The situation is depicted schematically
in \mbox{Fig. \ref{fig6}}.

Finally,
the above discussion
provides physical insight as to
why the ground state of the Gutzwiller-projected BCS Hamiltonian
is fundamentally different
from Anderson's RVB state
at half filling.
The difference originates from the singularity due to
the SDW instability which occurs at half filling.
This singularity cannot be captured
by applying the Gutzwiller projection to the BCS wavefunction
which, as constructed,
intrinsically lacks long-range antiferromagnetic order.
In fact, generally, no long-range order can be generated by applying
local operators to states without long-range order.
On the other hand,
long-range antiferromagnetic order is fully captured
in the ground state of
the Gutzwiller-projected BCS Hamiltonian,
as shown in the analytic derivation.
It has been shown in numerical studies that
the semi-classical N\'{e}el configuration has a finite weight
in the ground state of
the Gutzwiller-projected BCS Hamiltonian \cite{comment}.
Despite the difference at half filling, however,
we argue in Sec. \ref{conclusion} that,
sufficiently away from half filling,
the ground state of the Gutzwiller-projected BCS Hamiltonian
is, in fact, intimately related to the RVB state.

\subsection{Step (i)}
\label{step_i}

It is well known that
the nesting property of the Fermi surface at half filling
induces an instability toward
long-range antiferromagnetic order
(i.e., N\'{e}el order)
in the ground state of the Hubbard model.
Nesting is mathematically defined as follows:
\begin{eqnarray}
\epsilon_{\bf k} = -\epsilon_{{\bf k}+{\bf Q}} ,
\end{eqnarray}
where $\epsilon_{\bf k}= -2t(\cos{k_x}+\cos{k_y})$
is the kinetic energy due to hopping.
${\bf Q}=(\pi,\pi)$ is the nesting vector.
The strong-pairing BCS Hamiltonian
with $d$-wave pairing symmetry
has precisely the same nesting property
for the gap function:
\begin{eqnarray}
\tilde{\Delta}_{\bf k} = -\tilde{\Delta}_{{\bf k}+{\bf Q}} ,
\end{eqnarray}
where $\tilde{\Delta}_{\bf k} = 2\Delta_0(\cos{k_x}-\cos{k_y})$
is the gap function for $d$-wave pairing.
Therefore, in analogy with the Hubbard model,
the nesting property of $\tilde{\Delta}_{\bf k}$
generates a SDW instability at ($\pi,\pi$).
Note that, for both Hamiltonians,
the chemical potential is set to zero at half filling.
Away from half filling, the chemical potential becomes non-zero,
in which case the perfect nesting is ruined and
the SDW instability disappears.

In order to study spin orders,
it is convenient to re-write the on-site interaction Hamiltonian
in terms of spin operator,
${\bf S}_i=\frac{1}{2} c^{\dagger}_{ia} {\bf \sigma}_{ab} c_{ib}$:
\begin{eqnarray}
U\sum_i n_{i\uparrow}n_{i\downarrow}
= -\frac{2U}{3} \sum_i {\bf S}^2_i
+\frac{U}{6} \sum_i (n_{i\uparrow}+n_{i\downarrow})
\label{UtoS}
\end{eqnarray}
where the last term becomes constant
if the total number of electrons is fixed,
and will henceforth be suppressed
except when its consideration is necessary.

Keeping in mind that there is an intrinsic SDW instability
at half filling,
we decompose the spin operator into
the stationary and fluctuation part:
${\bf S}_i = \langle {\bf S}_i \rangle
+({\bf S}_i-\langle {\bf S}_i \rangle)$.
Then, by retaining all terms up to first order in
${\bf S}_i-\langle {\bf S}_i \rangle$,
one obtains the saddle-point Hamiltonian, ${\cal H}_{\Delta_0+U}$:
\begin{eqnarray}
{\cal H}_{\Delta_0+U} &=&
\Delta_0 \sum_{i}
(c^{\dagger}_{i\uparrow}c^{\dagger}_{i+\hat{x},\downarrow}
-c^{\dagger}_{i\downarrow}c^{\dagger}_{i+\hat{x},\uparrow}
+\textrm{H.c.})
\nonumber \\
&-& \Delta_0 \sum_{i}
(c^{\dagger}_{i\uparrow}c^{\dagger}_{i+\hat{y}\downarrow}
-c^{\dagger}_{i\downarrow}c^{\dagger}_{i+\hat{y},\uparrow}
+\textrm{H.c.})
\nonumber \\
&+& \frac{3}{8U} \sum_i {\bf\phi}^2_i
+\sum_i {\bf \phi}_i \cdot {\bf S}_i \;,
\end{eqnarray}
where the $d$-wave pairing symmetry is explicitly written
in real space
and ${\bf\phi}_i \equiv -\frac{4U}{3}
\langle {\bf S}_i \rangle_{\Delta_0+U}$
is the spin expectation value
for the ground state of ${\cal H}_{\Delta_0+U}$.
The remaining terms form the fluctuation Hamiltonian,
$\delta {\cal H}_{\Delta_0+U}$:
\begin{eqnarray}
\delta {\cal H}_{\Delta_0+U} &=&
-\frac{2U}{3} \sum_i
\left(
{\bf S}_i +\frac{3}{4U} {\bf\phi}_i
\right)^2 \;.
\label{BCS+U}
\end{eqnarray}
Similarly,
the Hubbard Hamiltonian, $H_{\textrm{Hub}}$,
can be decomposed into two parts:
\begin{eqnarray}
{\cal H}_{\textrm{Hub}} &=&
-t\sum_{\langle i,j \rangle}
(c^{\dagger}_{ia}c_{ja}+\textrm{H.c.})
\nonumber \\
&+& \frac{3}{8U} \sum_i {\bf\varphi}^2_i
+\sum_i {\bf\varphi}_i \cdot {\bf S}_i \;,
\nonumber \\
\delta {\cal H}_{\textrm{Hub}} &=&
-\frac{2U}{3} \sum_i
\left(
{\bf S}_i +\frac{3}{4U} {\bf\varphi}_i
\right)^2 \;,
\label{Hubbard}
\end{eqnarray}
where
${\bf\varphi}_i \equiv -\frac{4U}{3}
\langle {\bf S}_i \rangle_\textrm{Hub}$
is the spin expectation value for
the ground state of ${\cal H}_{\textrm{Hub}}$.

Since we are interested in the SDW singularity at ($\pi,\pi$),
we set both
${\bf\phi}_i = \phi_0 \cos{({\bf Q}\cdot{\bf r}_i)} \hat{z}$ and
${\bf\varphi}_i = \varphi_0 \cos{({\bf Q}\cdot{\bf r}_i)} \hat{z}$
with ${\bf Q}=(\pi,\pi)$.
Finally,
the saddle-point solution is completed by the determination of
the optimal value of $\phi_0$ ($\varphi_0$)
by minimizing the ground state energy of ${\cal H}_{\Delta_0+U}$
(${\cal H}_{\textrm{Hub}}$)
with respect to $\phi_0$ ($\varphi_0$).
Note that the optimization of the ground energy
amounts to the saddle-point condition
for the auxiliary field in the path integral formulation \cite{Fradkin};
hence the label ``the saddle-point Hamiltonian''.

\subsection{Step (ii)}
\label{step_ii}

It is convenient to re-write
${\cal H}_{\Delta_0+U}$ in momentum space:
\begin{eqnarray}
{\cal H}_{\Delta_0+U} &=&
\frac{3}{8U} \phi^2_0
+\int \frac{d^2 k}{(2\pi)^2} \tilde{\Delta}_{{\bf k}}
\left(
c^{\dagger}_{{\bf k}\uparrow}c^{\dagger}_{-{\bf k}\downarrow}
+\textrm{H.c.}
\right)
\nonumber \\
&+&\frac{\phi_0}{4} \int \frac{d^2 k}{(2\pi)^2}
\left(
c^{\dagger}_{{\bf k}+{\bf Q},a}
\sigma^{ab}_{z} c_{{\bf k},b}
+\textrm{H.c.}
\right),
\end{eqnarray}
where the Einstein sum rule is applied on the $\sigma$ indices.
Moreover, by defining a spinor field $\Psi_{\bf k}$,
\begin{eqnarray}
\Psi_{\bf k}
=
\left(
\begin{array}{c}
c_{{\bf k}\uparrow} \\
c_{{\bf k}+{\bf Q},\uparrow}\\
c^{\dagger}_{-{\bf k}\downarrow}\\
c^{\dagger}_{-{\bf k}-{\bf Q},\downarrow}
\end{array}
\right) \;,
\end{eqnarray}
one can re-write ${\cal H}_{\Delta_0+U}$
in a convenient 4$\times$4 form:
\begin{eqnarray}
{\cal H}_{\Delta_0+U} =
\frac{3}{8U} \phi^2_0
+\int_{\Omega} \frac{d^2 k}{(2\pi)^2}
\Psi^{\dagger}_{\bf k}
{\cal M}_{\bf k}
\Psi_{\bf k} \;,
\end{eqnarray}
where
\begin{equation}
{\cal M}_{\bf k} =
\left(
\begin{array}{cc}
\frac{\phi_0}{2}\sigma_x   &    \tilde{\Delta}_{\bf k}\sigma_z    \\
\tilde{\Delta}_{\bf k}\sigma_z     &    \frac{\phi_0}{2}\sigma_x
\end{array}
\right) \;.
\end{equation}
The momentum integration, $\int_{\Omega} d^2 k$,
covers half of the Brillouin zone,
which, for convenience, we choose to be
the area in $k$-space bounded by $k_y=k_x\pm\pi$ and $k_y=-k_x\pm\pi$.

While it is straightforward to obtain
eigenvalues of ${\cal M}_{\bf k}$,
it is instructive to diagonalize
${\cal M}_{\bf k}$ in two steps.
First, we apply the following unitary transformation
onto ${\cal M}_{\bf k}$:
\begin{eqnarray}
{\cal M}^{\prime}_{\bf k} &=&
 \left(
\begin{array}{cc}
{\cal U} &    0    \\
0        &    {\cal U}^{-1}
\end{array}
\right)
{\cal M}_{\bf k}
\left(
\begin{array}{cc}
{\cal U}^{-1} &    0    \\
0        &    {\cal U}
\end{array}
\right)
\end{eqnarray}
where
\begin{equation}
{\cal U} =
\frac{1}{\sqrt{2}}
\left(
\begin{array}{cc}
1 & 1 \\
-1 & 1
\end{array}
\right)
\end{equation}
which also defines $\gamma_{\pm}({\bf k})$:
\begin{equation}
\left(
\begin{array}{c}
\gamma_{+\uparrow}({\bf k}) \\
\gamma_{-\uparrow}({\bf k}) \\
\gamma^{\dagger}_{+\downarrow}(-{\bf k}) \\
\gamma^{\dagger}_{-\downarrow}(-{\bf k})
\end{array}
\right)
=
\left(
\begin{array}{cc}
{\cal U} &    0    \\
0        &    {\cal U}^{-1}
\end{array}
\right)
\Psi_{\bf k}.
\end{equation}
Second,
${\cal M}^{\prime}_{\bf k}$ is diagonalized
via Bogoliubov transformation:
\begin{eqnarray}
\left(
\begin{array}{c}
\alpha_{\pm\uparrow}({\bf k}) \\
\alpha^{\dagger}_{\pm\downarrow}(-{\bf k})
\end{array}
\right)
=
\left(
\begin{array}{cc}
u_{\pm}({\bf k}) & -v_{\pm}({\bf k}) \\
v_{\pm}({\bf k}) & u_{\pm}({\bf k})
\end{array}
\right)
\left(
\begin{array}{c}
\gamma_{\pm\uparrow}({\bf k}) \\
\gamma^{\dagger}_{\pm\downarrow}(-{\bf k})
\end{array}
\right)
\end{eqnarray}
where
\begin{eqnarray}
u^2_{\pm}({\bf k})-v^2_{\pm}({\bf k})&=&\pm\frac{\phi_0/2}{E_{\bf k}},
\nonumber \\
-2u_{\pm}({\bf k})v_{\pm}({\bf k})&=&
\pm\frac{\tilde{\Delta}_{\bf k}}{E_{\bf k}},
\label{uv}
\end{eqnarray}
and
\begin{eqnarray}
E_{\bf k}=\sqrt{\tilde{\Delta}^2_{\bf k}+(\phi_0/2)^2}.
\end{eqnarray}
Then, it can be shown that
the ground state is a vacuum of the Bogoliubov quasiparticles:
\begin{eqnarray}
\alpha_{\pm,\sigma}({\bf k}) |\psi^\textrm{gr}_{\Delta+U}\rangle = 0 ,
\end{eqnarray}
which is satisfied by the following wavefunction:
\begin{eqnarray}
|\psi^\textrm{gr}_{\Delta_0+U}\rangle &=&
\prod_{{\bf k}\in\Omega}
[u_+({\bf k}) +v_+({\bf k})
\gamma^{\dagger}_{+\uparrow}({\bf k})
\gamma^{\dagger}_{+\downarrow}(-{\bf k})]
\nonumber \\
&\times&
[u_-({\bf k}) +v_-({\bf k})
\gamma^{\dagger}_{-\uparrow}({\bf k})
\gamma^{\dagger}_{-\downarrow}(-{\bf k})] |0\rangle \;.
\label{psi_gr}
\end{eqnarray}

Moreover,
minimizing the ground state energy with respect to $\phi_0$
leads to the saddle-point equation for $\phi_0$:
\begin{eqnarray}
\frac{3}{2U} = \int_{\Omega} \frac{d^2 k}{(2\pi)^2}
\frac{1}{\sqrt{\tilde{\Delta}^2_{\bf k}+(\phi_0/2)^2}}.
\label{Eq_sp}
\end{eqnarray}
The saddle-point equation in \mbox{Eq. (\ref{Eq_sp})}
has a remarkable property that
the solution exists for arbitrary values of $U$.
Especially intriguing is that
the solution is singular in the small-$U$ limit:
$\ln{\phi_0} \propto -\Delta_0/U$ for $U \rightarrow 0$.
As a consequence,
there is long-range antiferromagnetic order (i.e., $\phi_0 \neq 0$)
even for an arbitrarily weak interaction $U$,
which is, in turn, adiabatically connected to
the long-range order in the large-$U$ limit.

In the large-$U$ limit,
$\phi_0$ is proportional to $U$: $\phi_0=2U/3$.
The excitation spectrum $E_{\bf k}$, therefore,
has a large energy gap proportional to $U$.
Consequently, in the large-$U$ limit,
the ground state, $|\psi^\textrm{gr}_{\Delta_0+U}\rangle$,
is completely separated from other excitations
as far as the saddle-point Hamiltonian,
${\cal H}_{\Delta_0+U}$, is concerned.
Furthermore, one can show that the situation is
exactly the same for the Hubbard model:
the ground state of ${\cal H}_\textrm{Hub}$ is
separated from other excitations by
an energy gap proportional to $U$ in the large-$U$ limit.
Low-energy fluctuations must, then, come from
the fluctuation Hamiltonians,
$\delta{\cal H}_{\Delta_0+U}$ and $\delta{\cal H}_\textrm{Hub}$.

Before we discuss the effect of fluctuations,
however, we analyze ${\cal H}_\textrm{Hub}$
to show that, in the large-$U$ limit,
the ground state of ${\cal H}_{\Delta_0+U}$
is identical to that of ${\cal H}_{\textrm{Hub}}$.
The saddle-point solution of the Hubbard model
was discovered long ago \cite{Fradkin,Dagotto};
the ground state of ${\cal H}_\textrm{Hub}$ is
a fully filled Fermi sea state of
the ``negative-energy-mode quasiparticles'', associated with $\beta_{-}$,
while it is a vacuum for the ``positive-energy-mode quasiparticles'',
associated with $\beta_{+}$.
The precise definition of $\beta_{\pm}$ is given by:
\begin{eqnarray}
\left(
\begin{array}{c}
\beta_{+\uparrow}({\bf k}) \\
\beta_{-\uparrow}({\bf k}) \\
\beta_{+\downarrow}({\bf k}) \\
\beta_{-\downarrow}({\bf k})
\end{array}
\right)
=\left(
\begin{array}{cccc}
\xi_k & -\eta_k & 0 & 0   \\
\eta_k & \xi_k  & 0 & 0   \\
0 & 0 & \xi_k & \eta_k \\
0 & 0 & -\eta_k  &  \xi_k
\end{array}
\right)
\left(
\begin{array}{c}
c_{{\bf k}\uparrow} \\
c_{{\bf k}+{\bf Q},\uparrow} \\
c_{{\bf k}\downarrow} \\
c_{{\bf k}+{\bf Q},\downarrow}
\end{array}
\right) ,
\end{eqnarray}
where
\begin{eqnarray}
\xi^2_k-\eta^2_k &=& \frac{\epsilon_{\bf k}}{{\cal E}_{\bf k}},
\nonumber \\
-2 \xi_k \eta_k &=& \frac{\varphi_0/2}{{\cal E}_{\bf k}},
\label{xieta}
\end{eqnarray}
and
\begin{eqnarray}
{\cal E}_{\bf k} =\sqrt{\epsilon^2_{\bf k}+(\varphi_0/2)^2} .
\end{eqnarray}
The ground state of ${\cal H}_\textrm{Hub}$ is, then, given by
\begin{eqnarray}
|\psi^\textrm{gr}_\textrm{Hub} \rangle =
\prod_{{\bf k} \in \Omega}
\beta^{\dagger}_{-\uparrow}({\bf k})
\beta^{\dagger}_{-\downarrow}(-{\bf k})
|0\rangle .
\label{grHub}
\end{eqnarray}

Similar to what was done
in the case of the strong-pairing BCS Hamiltonian,
minimizing the ground state energy leads to
the saddle-point equation for $\varphi_0$,
which is given by
\begin{eqnarray}
\frac{3}{2U} = \int_{\Omega} \frac{d^2 k}{(2\pi)^2}
\frac{1}{\sqrt{\epsilon^2_{\bf k}+(\varphi_0/2)^2}} .
\label{Eq_sp2}
\end{eqnarray}
It can be shown from \mbox{Eq. (\ref{Eq_sp2})}
that, in the large-$U$ limit,
$\varphi_0$ becomes proportional to $U$;
more precisely $\varphi_0 = 2U/3$, in which case
$\xi_k = -\eta_k = 1/\sqrt{2}$
[See \mbox{Eq. (\ref{xieta})}].
At the same time, for the strong-coupling BCS Hamiltonian,
$\phi_0$ is also proportional to $U$ in the large-$U$ limit
so that
$u_+({\bf k}) = 1$, $v_+({\bf k}) = 0$,
$u_-({\bf k}) = 0$, and $v_-({\bf k}) = 1$
[See \mbox{Eq. (\ref{uv})}],
which results in the following ground state:
\begin{eqnarray}
|\psi_0\rangle =
\prod_{{\bf k}\in\Omega}
\gamma^{\dagger}_{-\uparrow}({\bf k})
\gamma^{\dagger}_{-\downarrow}(-{\bf k})
|0\rangle .
\label{psi0}
\end{eqnarray}
As one can see,
$|\psi_0\rangle$ is identical to
the ground state of ${\cal H}_{\textrm{Hub}}$
in \mbox{Eq. (\ref{grHub})}
if the large-$U$ limit is taken
(Note that $\beta_{-}({\bf k}) = \gamma_{-}({\bf k})$
in the large-$U$ limit).

It should not come as a surprise that
$\langle {\bf S}_i \rangle$
is independent of parameters (such as $U/t$ or $U/\Delta_0$)
since, in the large-$U$ limit,
the ground state, $|\psi_0\rangle$, itself
is uniquely determined without
the dependence on $U/t$ or $U/\Delta_0$.
Remember that, in the large-$U$ limit,
the Hubbard model becomes the Heisenberg model
which has a single scale factor $J$. Therefore,
the ground state (actually, any eigenstate)
of the Heisenberg model is parameter free.
The situation is similar for
the strong-pairing Gutzillwer-projected BCS Hamiltonian,
$H^\textrm{G}_{\Delta}$.

So, at least superficially, it seems that
taking the large-$U$ limit
entails simply
making $U$ infinite
and ignoring any effect of $t$ or $\Delta_0$.
However, this is not correct
because  there is a reduction in the ground state energy
due to finite $\Delta_0$ in $H_{\Delta_0+U}$.
The ground state energy
of $|\psi^\textrm{gr}_{\Delta_0+U}\rangle$
has the following form:
\begin{eqnarray}
\frac{E^\textrm{gr}_{\Delta_0+U}}{N_e} = \frac{3}{8U} \phi^2_0
-2 \int_{\Omega} \frac{d^2 k}{(2 \pi)^2}
\sqrt{\tilde{\Delta}^2_{\bf k}+(\phi_0/2)^2}
+\frac{U}{6} ,
\end{eqnarray}
where
the last term $U/6$ comes from the constant term
in Eq. (\ref{UtoS}).
Then,
it can be shown that, in the large-$U$ limit,
\begin{eqnarray}
\frac{E^\textrm{gr}_{\Delta_0+U}}{N_e} &\simeq&
-\frac{2}{\phi_0} \int_{\Omega} \frac{d^2 k}{(2 \pi)^2}
\tilde{\Delta}^2_{\bf k}
\nonumber \\
&=& -\frac{3\Delta_0^2}{U} \int_{\Omega} \frac{d^2 k}{(2 \pi)^2}
(\cos{k_x}-\cos{k_y})^2
\nonumber \\
&=& -\frac{3\Delta_0^2}{U},
\label{EgrDeltaU}
\end{eqnarray}
where we use the fact that $\phi_0 =2U/3$ in the large-$U$ limit.
As one can see, for finite $\Delta_0$,
there is a reduction in the ground state energy,
which is proportional to $\Delta_0^2/U$.
As long as $\Delta_0$ is not zero,
spin configurations are not random in
the low-energy Hilbert space.
This is the difference between the large-$U$ limit
and strictly infinite $U$.

The situation is similar for the Hubbard model.
The ground state energy
of $|\psi^\textrm{gr}_\textrm{Hub}\rangle$
is given by:
\begin{eqnarray}
\frac{E^\textrm{gr}_\textrm{Hub}}{N_e} =
\frac{3}{8U} \varphi^2_0
-2 \int_{\Omega} \frac{d^2 k}{(2 \pi)^2}
\sqrt{\epsilon^2_{\bf k}+(\varphi_0/2)^2}
+\frac{U}{6} ,
\end{eqnarray}
which, in the large-$U$ limit, becomes
\begin{eqnarray}
\frac{E^\textrm{gr}_\textrm{Hub}}{N_e} &\simeq&
-\frac{2}{\varphi_0} \int_{\Omega} \frac{d^2 k}{(2 \pi)^2}
\epsilon^2_{\bf k}
\nonumber \\
&=& -\frac{3 t^2}{U} \int_{\Omega} \frac{d^2 k}{(2 \pi)^2}
(\cos{k_x}+\cos{k_y})^2
\nonumber \\
&=& -\frac{3 t^2}{U}.
\label{EgrHub}
\end{eqnarray}
Therefore, again,
spin configurations of the low-energy states
are not random.

\subsection{Step (iii)}
\label{step_iii}

We have shown that
the excitation spectra of
${\cal H}_{\Delta_0+U}$ and ${\cal H}_\textrm{Hub}$
are gapped with a very large energy gap of ${\cal O}(U)$
in the large-$U$ limit.
However, the true low-energy excitation should be gapless
at half filling, as required by Goldstone's theorem
in two dimension where the spin rotation symmetry is broken
by non-zero $\langle{\bf S}_i\rangle$.
This dilemma is an artifact
of the restriction of our attention to
only a part of the full Hamiltonian.

The saddle-point ground state, $|\psi_0\rangle$,
is separated from other excitations
of the saddle-point Hamiltonian.
However, when the full Hamiltonian
(including the saddle-point and fluctuation part)
is considered,
there are many other states which have exactly
the same energy as $|\psi_0\rangle$;
these states are connected to
$|\psi_0\rangle$ via rigid spin rotations.
We denote these states as
$|\psi_i\rangle \equiv {\bf R}_i |\psi_0\rangle$
with $\{{\bf R}_i\}$ being spin rotation operators.
The energy of $\{|\psi_i\rangle\}$ is the same
for an arbitrary spin rotation because
\begin{eqnarray}
\langle \psi_i | H | \psi_i \rangle
=\langle \psi_0 | {\bf R}^{\dagger}_i H
{\bf R}_i | \psi_0 \rangle
=\langle \psi_0 | H | \psi_0 \rangle ,
\end{eqnarray}
where $H$, either $H_{\Delta+U}$ or $H_\textrm{Hub}$,
is invariant under spin rotation.
In the large-$U$ limit, therefore,
the low-energy Hilbert space
is composed only of $\{|\psi_i\rangle\}$.
Note that, although linearly independent,
the basis states in $\{|\psi_i\rangle\}$
are not orthogonal to each other;
$\{|\psi_i\rangle\}$ forms an over-complete basis
set, which is conceptually very similar
to the coherent-state basis set for a spin representation \cite{Auerbach}.

Our task now is to evaluate
matrix elements of the full Hamiltonian
within the above low-energy Hilbert space.
To this end, we first investigate the saddle-point equation.
Eq. (\ref{Eq_sp2}) is very similar to Eq. (\ref{Eq_sp}).
In fact,
when all quantities are properly scaled,
Eq. (\ref{Eq_sp2})
is exactly the same equation as Eq. (\ref{Eq_sp})
since the momentum, {\bf k}, is a dummy integration parameter
and Eq. (\ref{Eq_sp2}) can be obtained from Eq. (\ref{Eq_sp})
by translating ${\bf k}$ by ($\pi,0$).
A constant translation in momentum
does not affect the integral in the saddle-point equation.
To be specific,
if the solution of Eq. (\ref{Eq_sp2}), $\tilde{\varphi}_0/t$,
is a function of $U/t$:
\begin{eqnarray}
\tilde{\varphi}_0/t = f(U/t) ,
\end{eqnarray}
then the solution of Eq. (\ref{Eq_sp}), $\tilde{\phi}_0/\Delta_0$,
is related to $U/\Delta_0$ by the same function $f$:
\begin{eqnarray}
\tilde{\phi}_0/\Delta_0 = f(U/\Delta_0) .
\end{eqnarray}
The precise functional form of $f(x)$ is
not important for general $x$, but
it is useful to know that $f(x) = 2x/3$ when $x \gg 1$.

Now,
considering the definition of
$\delta{\cal H}_{\Delta_0+U}$ and $\delta{\cal H}_\textrm{Hub}$,
it is not too difficult to show that
effects of both fluctuation Hamiltonians become equivalent
in the large-$U$ limit
since
(i) $\tilde{\phi}_0 = \tilde{\varphi}_0$ for general $U$,
if $t$ is set equal to $\Delta_0$, and
(ii) low-energy Hilbert spaces for both models
reduce to the same Hilbert space, $\{|\psi_i\rangle\}$,
in the large-$U$ limit.
So, in essence, it has been shown that, in the large-$U$ limit,
all matrix elements for both fluctuation Hamiltonians
are the same in the low-energy Hilbert space.
It is very important to note that
we are interested in the the lowest-order, non-zero effect of
finite $\Delta_0$ and $t$ in the limit of large $U$.
Since, in the large-$U$ limit,
eigenstates themselves should be parameter free,
lowest-order effects of finite finite $\Delta_0$ and $t$
must emerge through their effects on
$\phi_0$ and $\varphi_0$, respectively.

We now establish that matrix elements of
both saddle-point Hamiltonians,
${\cal H}_{\Delta_0+U}$ and ${\cal H}_\textrm{Hub}$,
are also identical.
We begin by considering
two arbitrary states from the low-energy Hilbert space
of the BCS Hamiltonian with finite $U$,
$|\tilde{\psi}_j\rangle= {\bf R}_j |\psi^\textrm{gr}_{\Delta_0+U}\rangle$
and
$|\tilde{\psi}_k\rangle= {\bf R}_k |\psi^\textrm{gr}_{\Delta_0+U}\rangle$.
The matrix element of ${\cal H}_{\Delta_0+U}$
between these two states is, then,
\begin{eqnarray}
&&\Big\langle\tilde{\psi}_j \Big|
{\cal H}_{\Delta_0+U}
\Big| \tilde{\psi}_k \Big\rangle
\nonumber \\
&=&
\Big\langle\tilde{\psi}_j \Big|
H_{\Delta_0}
+\frac{3}{8U} \sum_i {\bf\phi}^2_i
+\sum_i {\bf\phi}_i \cdot {\bf S}_i
\Big| \tilde{\psi}_k \Big\rangle
\nonumber \\
&=& E^\textrm{gr}_{\Delta_0+U}
\Big\langle\psi^\textrm{gr}_{\Delta_0+U}\Big|
{\bf R}^{\dagger}_j {\bf R}_k
\Big|\psi^\textrm{gr}_{\Delta_0+U}\Big\rangle
\nonumber \\
&+&
\Big\langle\psi^\textrm{gr}_{\Delta_0+U}\Big|
{\bf R}^{\dagger}_j
\Big(\sum_i{\bf\phi}_i \cdot {\bf S}_i\Big)
{\bf R}_k
\Big|\psi^\textrm{gr}_{\Delta_0+U}\Big\rangle
\nonumber \\
&-&\frac{1}{2}
\Big\langle\psi^\textrm{gr}_{\Delta_0+U}\Big|
\Big\{
{\bf R}^{\dagger}_j {\bf R}_k ,
\sum_i{\bf\phi}_i \cdot {\bf S}_i
\Big\}
\Big|\psi^\textrm{gr}_{\Delta_0+U}\Big\rangle ,
\label{me1}
\end{eqnarray}
where
$E^\textrm{gr}_{\Delta_0+U} = -3\Delta_0^2/U$
and $|\psi^\textrm{gr}_{\Delta_0+U}\rangle = |\psi_0\rangle$
in the large-$U$ limit.
Note that the invariance of $H_{\Delta_0}$ and ${\bf\phi}^2$
under spin rotations is used.
Similarly, in the large-$U$ limit,
the matrix element of ${\cal H}_\textrm{Hub}$
between the above states
is as follows:
\begin{eqnarray}
&&\Big\langle\psi_j \Big|
{\cal H}_\textrm{Hub}
\Big| \psi_k \Big\rangle
\nonumber \\
&=& E^\textrm{gr}_\textrm{Hub}
\Big\langle\psi_0\Big|
{\bf R}^{\dagger}_j {\bf R}_k
\Big|\psi_0\Big\rangle
\nonumber \\
&+&
\Big\langle\psi_0\Big|
{\bf R}^{\dagger}_j
\Big(\sum_i{\bf\varphi}_i \cdot {\bf S}_i\Big)
{\bf R}_k
\Big|\psi_0\Big\rangle
\nonumber \\
&-&\frac{1}{2}
\Big\langle\psi_0\Big|
\Big\{
{\bf R}^{\dagger}_j {\bf R}_k ,
\sum_i{\bf\varphi}_i \cdot {\bf S}_i
\Big\}
\Big|\psi_0\Big\rangle ,
\label{me2}
\end{eqnarray}
where $E^\textrm{gr}_\textrm{Hub} = -3t^2/U$
in the large-$U$ limit.
Therefore,
it is clear from a comparison of Eq. (\ref{me1}) and (\ref{me2})
that the matrix elements are identical,
provided that the large-$U$ limit is taken
while $t$ is set equal to $\Delta_0$.
Note that ${\bf\phi}_i = {\bf\varphi}_i$
for general $U$ if $t=\Delta_0$.


\subsection{Step (iv)}
\label{step_iv}

We have learned two lessons
from Step (ii) and (iii).
First, the ground states of the saddle-point Hamiltonians,
${\cal H}_\textrm{Hub}$ and ${\cal H}_{\Delta_0+U}$,
are identical in the large-$U$ limit,
and therefore
low-energy Hilbert spaces
(which are obtained by applying rigid spin rotations
to the saddle-point ground states)
are also identical in this limit.
Second, in the low-energy Hilbert space,
matrix elements for the fluctuation Hamiltonians
are identical for
both the Hubbard model and the strong-pairing BCS Hamiltonian
with the same being true for the two saddle-point Hamiltonians.
Therefore, we conclude that
the strong-pairing Gutzwiller-projected BCS Hamiltonian
at half filling
is equivalent to the 2D antiferromagnetic Heisenberg model.

\subsection{Relation to the projected BCS Hamiltonian
with extended $s$-wave pairing: a corollary}

In this section,
we would like to discuss an interesting corollary
of the analytic derivation given in previous sections.
Our derivation is actually valid also for
the strong-pairing BCS Hamiltonian
with an extended $s$-wave pairing:
$\Delta_{\bf k}/\Delta
=2(\cos{k_x}+\cos{k_y})$.
This is due to the fact that
the saddle-point equation for extended $s$-wave pairing
is the same as that of $d$-wave pairing and, consequently,
all matrix elements remain the same as obtained for $d$-wave case.
Note that extended $s$-wave pairing
is induced by nearest-neighbor pairing amplitudes
with the same sign for the $x$ and $y$ directions
in contrast to the opposite signs in $d$-wave pairing.
However, despite this important distinction,
it is proven as a corollary to our $d$-wave result that,
at half filling,
the strong-pairing Gutzwiller-projected BCS Hamiltonian
with extended $s$-wave pairing
is also equivalent to the antiferromagnetic Heisenberg model.

\begin{figure}
\includegraphics[width=2in,angle=-90]{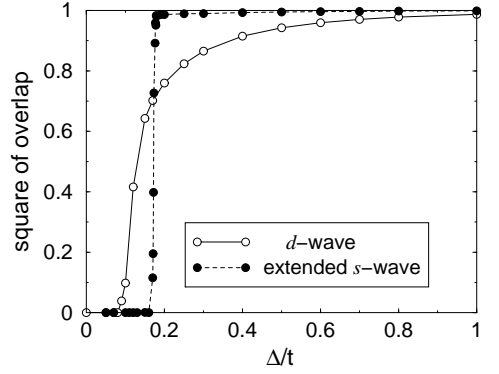}
\caption{Square of the overlap between the ground state
of the antiferromagnetic Heisenberg model, and
those of the Gutzwiller-projected BCS Hamiltonian
with $\Delta_{\bf k}=2\Delta (\cos{k_x}+\cos{k_y})$
[extended $s$-wave pairing]
and $2\Delta(\cos{k_x}-\cos{k_y})$ [$d$-wave pairing].
\label{fig7}}
\end{figure}

To test this assertion numerically,
we compute the wavefunction overlap between
the ground states of the Heisenberg model and
the strong-pairing Gutzwiller-projected BCS Hamiltonian
with extended $s$-wave pairing.
As before,
we analyze the 4$\times$4 square lattice system
via exact diagonalization.
To be specific,
the strong-pairing Gutzwiller-projected BCS Hamiltonian
with extended $s$-wave pairing is given by:
\begin{eqnarray}
H^\textrm{G}_{s\textrm{BCS}}=
\hat{{\cal P}}_G
( H_t + H_{s\Delta})
\hat{{\cal P}}_G ,
\end{eqnarray}
where
\begin{eqnarray}
H_{s\Delta} &=&
\Delta \sum_i
(c^{\dagger}_{i\uparrow}c^{\dagger}_{i+\hat{x},\downarrow}
-c^{\dagger}_{i\downarrow}c^{\dagger}_{i+\hat{x},\uparrow}
+\textrm{H.c.})
\nonumber \\
&+&\Delta \sum_i
(c^{\dagger}_{i\uparrow}c^{\dagger}_{i+\hat{y},\downarrow}
-c^{\dagger}_{i\downarrow}c^{\dagger}_{i+\hat{y},\uparrow}
+\textrm{H.c.} ) .
\end{eqnarray}
\mbox{Fig. \ref{fig7}} shows that,
for sufficiently large $\Delta/t$,
the overlap indeed becomes essentially unity for both
$\Delta_{\bf k} = 2 \Delta (\cos{k_x}+\cos{k_y})$
[extended $s$-wave pairing]
and $2\Delta(\cos{k_x}-\cos{k_y})$ [$d$-wave pairing].

\begin{figure}
\includegraphics[width=2.2in]{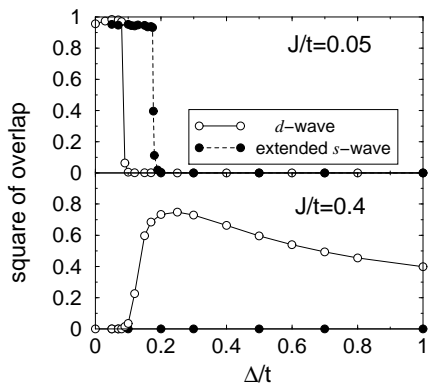}
\caption{Square of the overlap for the 2-hole case.
Plotted is the square of the overlap between the ground state
of the $t-J$ model and
those of the Gutzwiller-projected BCS Hamiltonian
with $\Delta_{\bf k}=2\Delta (\cos{k_x}+\cos{k_y})$
[extended $s$-wave pairing]
and $2\Delta(\cos{k_x}-\cos{k_y})$ [$d$-wave pairing].
\label{fig8}}
\end{figure}
As before in the $d$-wave case,
it is also interesting to examine the doped regimes.
While there is no difference
between extended $s$-wave and $d$-wave pairing at half filling,
the $t$-$J$ model prefers $d$-wave pairing at finite doping
(at least in the ${\bf k}=(0,0)$ channel).
\mbox{Fig. \ref{fig8}} shows
the wavefunction overlap for the $4\times4$ system with 2 holes.
Except for the trivial parameter regime of very small $J/t$
(top panel in Fig. \ref{fig8}),
the overlap essentially vanishes
for extended $s$-wave pairing
in contrast to the high overlap
for the case of $d$-wave pairing
(bottom panel in Fig. \ref{fig8}).


\section{Conclusion}
\label{conclusion}

We conclude by discussing the physical implications
of the close connection between
the Gutzwiller-projected BCS Hamiltonian and
the $t$-$J$ model.
There are practical implications as well as fundamental ones.
Fundamentally,
the close connection between
the Gutzwiller-projected BCS Hamiltonian and the $t$-$J$ model
provides evidence for
the existence of superconductivity in the $t$-$J$ model,
which, in turn, may suggest that
it is not only possible, but rather natural that
the pairing in high $T_C$ superconductors
is caused purely by electronic correlations.

Now that we have reason to believe that
the antiferromagnetic interaction is
closely related to electron-electron pairing,
the next natural question is
how this connection can be used
for the quantitative understanding of experiments.
While it is in principle straightforward to apply
exact diagonalization to
the Gutzwiller-projected BCS Hamiltonian,
exactly solving the Gutzwiller-projected BCS Hamiltonian
is as complicated as
solving the $t$-$J$ model in the first place.
The true practical advantage of analyzing
the Gutzwiller-projected BCS Hamiltonian, however,
lies in the fact that
the Gutzwiller-projected BCS Hamiltonian
is much more amenable to the ansatz wavefunction approach
than the $t$-$J$ model.

Ansatz wavefunction approaches have been very successful
in various strongly correlated problems
in condensed matter physics.
Two very salient examples are
the fractional quantum Hall effect (FQHE)
and the liquid Helium.
Despite their apparent difference in the physical context,
there is actually a rather profound commonality
between the above, two examples.
In both cases,
dominant short-range correlations are first
separated from long-range correlations.
Then, effects of the strong short-range correlations
are captured by the attachment of the Jastrow factor.
The specific functional form of the Jastrow factor, of course,
depends on the nature of the problem at hand.
After short-range correlations are taken into account
via the Jastrow factor,
it is assumed that
residual long-range correlations are relatively much weaker
so that the rest of the ansatz wavefunction describes
essentially non-interacting quasiparticles.

Specifically,
the ansatz wavefunction for the FQHE
is given by the composite fermion (CF) wavefunction \cite{Jain}.
In essence,
the CF wavefunction is a product of the Jastrow factor,
$\prod_{i<j} (z_i - z_j)^{2p}$
with $p$ an integer and $z_j=x_j + i y_j$,
and a non-interacting fermionic wavefunction of
new quasiparticles known as composite fermions.
It is known
that the above functional form of
the Jastrow factor is most effective
in minimizing the Coulomb repulsion between electrons
in the lowest Landau level;
the wavefunction overlap
between the CF wavefunction and the exact ground state
is practically unity for all available
finite-system studies \cite{DasSarmaPinczuk}.

Another example is liquid Helium.
The microscopic wavefunction
for normal state $^3$He
is written in terms of an algebraic product of the Jastrow factor
and the Slater determinant of plane-wave states \cite{Mahan}:
\begin{equation}
\Psi_{^3\textrm{He}} =
\exp{\Big[-\sum_{i<j} u({\bf r}_i-{\bf r}_j)\Big]}
\textrm{Det}\left|e^{i{\bf k}\cdot{\bf r}}\right|
\end{equation}
where $u({\bf r})$ is determined by
the interaction between He atoms:
typically, $u({\bf r}) \propto 1/r^5$.
While the Slater determinant provides
the necessary antisymmetrization
as well as the low-energy physics,
the Jastrow factor captures
dominant short-range correlations.
Similarly, the ansatz wavefunction for superfluid $^4$He
is given by:
\begin{equation}
\Psi_{^4\textrm{He}} =
\exp{\Big[-\sum_{i<j} u'({\bf r}_i-{\bf r}_j)\Big]}
\end{equation}
where $u'({\bf r})$ is basically identical
$u({\bf r})$ except for slight changes
due to the mass difference between $^3$He and $^4$He.
Since the ground state wavefunction for non-interacting bosons
is just a constant,
the ansatz wavefunction for superfluid $^4$He
can be also viewed as
the product of the Jastrow factor and
a wavefunction for non-interacting bosons.

From the discussion so far,
it seems promising that
a range of strongly correlated problems
can be attacked via the Jastrow-factor approach,
provided that residual, long-range correlations are weak
so that, to a good approximation,
they can be treated as
those of non-interacting quasi-particles.
One criterion for the weakness of residual interaction may be
whether or not the residual interaction can cause
an instability toward a new phase.
In other words,
as long as the ground state for the original interaction
(causing the strong short-range correlation)
is not completely different from
that of weakly-interacting (emergent) quasi-particles,
the Jastrow-factor approach can provide
a good ansatz wavefunction.

The situation for the $t$-$J$ model, or the Hubbard model in the
large-$U$ limit, is quite intriguing. First, note that the
Gutzwiller projection can be actually regarded as the Jastrow
factor imposing the no-double-occupancy constraint that occurs
because of the strong on-site repulsion in the large-$U$ limit.
Keeping in mind that the $t$-$J$ model is the Hubbard model in the
large-$U$ limit, it may be conjectured that the ground state of
the $t$-$J$ model is described well by the Gutzwiller-projected
Fermi-sea state. Unfortunately, however, this is not true at half
filling. We have learned in Sec. \ref{step_ii} that, at half
filling, even the infinitesimally weak interaction, $U$, can cause
a spin-density-wave (SDW) instability leading to long-range
antiferromagnetic order, i.e., N\'{e}el order. Therefore, at least
close to half filling where N\'{e}el order persists, the
Gutzwiller-projected Fermi-sea state cannot be a good ansatz
wavefunction.

On the other hand, away from half filling, the SDW instability
disappears because perfect nesting is ruined with non-zero doping,
as shown in Sec. \ref{step_i}. One may assume now that the
Gutzwiller-projected Fermi-sea state can be a good ansatz
wavefunction for finite, non-zero doping. This is, however, not
true either because there is another instability caused by
electron-electron pairing. It is, in some sense, the goal of this
paper to show that the $t$-$J$ model contains a pairing
instability. Motivated by the fact that the Gutzwiller projection
can play a role of the Jastrow factor for the Gutzwiller-projected
BCS Hamiltonian, we now show that the Gutzwiller-projected BCS
wavefunction (i.e., the RVB state) is, in fact, a good ansatz
wavefunction for the Gutzwiller-projected BCS Hamiltonian at
finite, non-zero doping, which, in turn, leads to the eventual
connection between the ground state of the $t-J$ model and the
projected BCS wavefunction.

\begin{figure}
\includegraphics[width=2.2in]{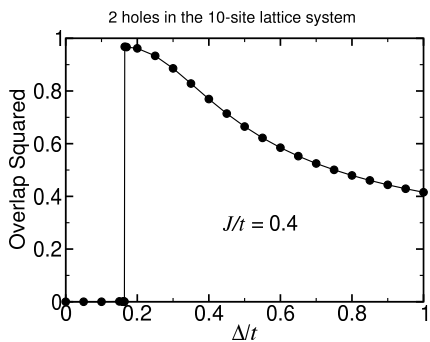}
\caption{Square of the overlap for the case of 2 holes in the
$\sqrt{10}\times\sqrt{10}$ system. Plotted is the square of the
overlap between the ground state of the $t-J$ model and those of
the Gutzwiller-projected BCS Hamiltonian with
$2\Delta(\cos{k_x}-\cos{k_y})$ [$d$-wave pairing]. The
discontinuity in the square of the overlap is due to the sudden
symmetry change in the ground state. \label{fig9}}
\end{figure}
While there have been studies showing that, for finite doping, the
Gutzwiller-projected BCS wavefunction yields good agreement with
experiments \cite{PRT} as well as with some numerical studies
\cite{Sorella}, we provide a more direct piece of evidence.
Hasegawa and Poilblanc \cite{HP} have shown that the overlap
between the projected BCS wavefunction and the ground state of the
$t-J$ model is high ($\sim90\%$) for the case of 2 holes in the
$\sqrt{10}\times\sqrt{10}$ lattice system. In order to make a
direct comparison, we have computed the overlap between the ground
state of the projected BCS Hamiltonian and that of the $t-J$ model
in the same system; we find that the optimal overlap for the
ground state of the projected BCS Hamiltonian is more than $98\%$
for the same $\sqrt{10}\times\sqrt{10}$ lattice system, as shown
in Fig. \ref{fig9}. Therefore, it is shown (at least, in finite
system studies) that the projected BCS wavefunction is a good
ansatz wavefunction for the projected BCS Hamiltonian, as expected
from the Jastrow-factor approach.

Regarding the work of Hasegawa and Poilblanc \cite{HP}, it is
interesting to note that they also computed the overlap between
the projected BCS wavefunction and the ground state of the $t-J$
model in the $4\times4$ lattice system and they found that it was
very low ($\sim4\%$). This sudden drop in the overlap is a
finite-system-size artifact due to the fact that the half-filled
Fermi sea state (technically speaking, its chemical potential) is
ill-defined in the $L\times L$ system with $L$ an even integer
(e.g., the $4\times4$ system) , while there is no such problem in
the $L\times L + 1$ system with $L$ an odd integer (e.g., the
$\sqrt{10}\times\sqrt{10}$ system when $L=3$). It is important to
note that the ground state of the projected BCS Hamiltonian in our
approach does not suffer from the above problem regardless of site
numbers.

The close relationship between the projected BCS wavefunction and
the ground state of the projected BCS Hamiltonian at moderate
doping is rather important since it provides additional support
for the existence of superconductivity in the projected BCS
Hamiltonian and, eventually, for that of the superconductivity in
the $t-J$ model through the connection between the projected BCS
Hamiltonian and the $t-J$ model; large-scale Monte Carlo
simulation studies have shown \cite{PRT,Sorella} that the
projected BCS wavefunction has the long-range pairing correlation
in the thermodynamic limit. In this perspective, the goal of our
work in this paper is to show why the projected BCS wavefunction
can be a good ansatz wavefunction for the $t-J$ model at moderate
doping, while it is not true at half filling. The
interrelationship between the ground state of the $t-J$ model,
that of the Gutzwiller-projected BCS Hamiltonian, and the RVB
state is schematically shown in Table \ref{table1}.
\begin{table}
\centering
\begin{tabular}{c|c}
\hline\hline
half filling & $\psi_{tJ} = \psi^\textrm{G}_\textrm{BCS} \ne \psi_\text{RVB}$ \\
\hline moderate doping & $\psi_{tJ} \simeq
\psi^\textrm{G}_\textrm{BCS} \simeq \psi_\text{RVB}$ \\
\hline\hline
\end{tabular}
\caption{Interrelationship between the ground state of the $t-J$
model, $\psi_{tJ}$, the ground state of the Gutzwiller-projected
BCS Hamiltonian, $\psi^\textrm{G}_\textrm{BCS}$, and the RVB
state, $\psi_\textrm{RVB}$. } \label{table1}
\end{table}

It is interesting to note that the recently proposed state for the
Gossamer superconductor \cite{Gossamer1,Gossamer2} is actually
the BCS wavefunction with a partial Gutzwiller projection allowing
the double occupancy with suppression weights:
\begin{eqnarray}
\hat{\cal P}^\textrm{partial}_\textrm{G} &\equiv& \prod_i
(1-\alpha n_{i\uparrow}n_{i\downarrow}) \nonumber \\
&=& e^{-\beta \sum_i n_{i\uparrow}n_{i\downarrow}} ,
\end{eqnarray}
where $e^{-\beta}=1-\alpha$. The above expression suggests a very
natural interpretation that the partial Gutzwiller projection is,
in fact, also a specific form of the usual Jastrow factor.
Incidentally, the Fermi sea state with the partial Gutzwiller
projection, $\hat{\cal P}^\textrm{partial}_\textrm{G}$, was
studied previously in the context of the metal-insulator
transition in the two-dimensional Hubbard model
\cite{BrinkmanRice}.

Finally, it is also interesting to note that, in one dimension
where there is no instability for either spin-density-wave or
pairing, the exact ground state of the antiferromagnetic
Heisenberg model (i.e., Bethe ansatz solution \cite{Bethe}) is
actually very closely related to the Gutzwiller-projected
Fermi-sea state \cite{GJR,KHF,GV}, as expected from the
Jastrow-factor argument.

\section{Acknowledgement}
\label{acknowledgement}
The author is very grateful to S. Das Sarma
for his support throughout this work, and
to J. K. Jain for his insightful comment on
the difference between the projected Hamiltonian
and the projected ground state.
Valuable discussions with
V. M. Galitski, D. J. Priour, Jr. and V. W. Scarola
are also greatly appreciated.
The author also acknowledges
very helpful correspondences with P. A. Lee and F. C. Zhang.
This work was supported by ARDA and NSF.



\begin{thebibliography}{}
\bibitem{comment} We stress, however, that
the semi-classical N\'{e}el configuration alone
has a very small overlap with the exact ground state.
The overlap squared is roughly
$16 \%$ in the 4$\times$4 system.
\end{thebibliography}
\end{document}